\documentclass[apj]{emulateapj}

\usepackage{natbib}
\usepackage{color}
\usepackage{graphicx}
\bibpunct{(}{)}{;}{a}{}{,}

\newcommand\paperi{Paper~I}
\newcommand\nxmps{{196}}
\newcommand\nxmpsdos{{332}}
\newcommand\ndrts{{788677}}
\newcommand\ncandidates{1281}
\newcommand\duplicated{11}

\begin{document}

   \title{
Search for Extremely Metal-poor Galaxies in the Sloan Digital Sky Survey (II): \\
high electron temperature objects
}
   \author{
                J.~S\'anchez~Almeida\altaffilmark{1,2},
                E. P\'erez-Montero\altaffilmark{3},
                A. B. Morales-Luis\altaffilmark{1,2},
        	C.~Mu\~noz-Tu\~n\'on\altaffilmark{1,2}, 
                R. Garc\'\i a-Benito\altaffilmark{3},
                S. E. Nuza\altaffilmark{4},
                and
                F.~S. Kitaura\altaffilmark{4},
          }
\altaffiltext{1}{Instituto de Astrof\'\i sica de Canarias, E-38205 La Laguna,
Tenerife, Spain}
\altaffiltext{2}{Departamento de Astrof\'\i sica, Universidad de La Laguna,
Tenerife, Spain}
\altaffiltext{3}{Instituto de Astrof\'\i sica de Andaluc\'\i a,  CSIC, PO box 3004, 
E-18080 Granada, Spain}
\altaffiltext{4}{Leibniz-Institut f\"ur Astrophysik Potsdam (AIP), An der Sternwarte 16, D-14482 Potsdam, Germany}
\email{jos@iac.es}
\begin{abstract}
Extremely metal-poor (XMP) galaxies are defined to have gas-phase metallicity smaller 
than a tenth of the solar value ($12+\log[{\rm O/H}] < 7.69$). 
They are uncommon, chemically and possibly dynamically 
primitive, with physical conditions characteristic of 
earlier phases of the  Universe. We search for new 
XMPs in the Sloan Digital Sky Survey (SDSS) in a work that complements Paper I. 
This time high electron temperature objects are selected; 
since metals are a main coolant of the gas, 
metal-poor objects contain high-temperature gas. Using the algorithm 
k-means, we classify \ndrts\ spectra to select \ncandidates\ galaxies 
having particularly intense [OIII]$\lambda$4363
with respect to [OIII]$\lambda$5007, which is a proxy for high 
electron temperature.  The metallicity of these candidates was computed using a hybrid technique consistent
with the direct method, rendering \nxmps\ XMPs.  
A less restrictive noise constraint provides
 a larger set with \nxmpsdos\ candidates. Both lists are provided in electronic format.
The selected XMP sample have mean stellar mass around $10^8\,{\rm M}_\odot$,
with dust-mass $\sim 10^3{\rm M}_\odot$ for typical star-forming regions.
In agreement with previous findings, XMPs show a tendency to be
tadpole-like or cometary. Their underlying stellar continuum corresponds to a 
fairly young stellar population ($< 1\,{\rm Gyr}$), although young and aged stellar 
populations co-exists at the low-metallicity starbursts.
About 10\,\%\ of the XMPs shows large N/O. Based on their location in constrained 
cosmological numerical  simulations, XMPs have a strong tendency to appear in voids 
and to avoid galaxy clusters. The puzzling 2\,\%-solar low-metallicity 
threshold exhibited  by XMPs remains.
\end{abstract}
%
%
\slugcomment{2nd version submitted to ApJ}

\shorttitle{Extremely metal poor galaxies of high electron temperature}



\section{Introduction}\label{introduction}

Extremely metal-poor (XMP) galaxies are defined to have a gas-phase metallicity 
ten times smaller than the sun \citep[e.g.,][]{2000A&ARv..10....1K}. 
Even though the exact threshold is arbitrary,
this definition selects chemical unevolved galaxies thus providing a gateway to study 
physical processes in conditions characteristic of the early 
Universe, which are now unusual. XMPs have been used to determine
the primordial He abundance produced during  the 
big bang \citep{2010IAUS..268...91P,2010JCAP...04..029S},
to study the star-formation in conditions of low metallicity 
\citep{2014Natur.514..335S,Filho+15b}, 
to infer the formation of dust in the early Universe \citep{2014Natur.505..186F}, 
to analyze properties of a primitive inter-stellar medium \citep{2007ApJ...665.1115I}, 
or to constrain the properties of the first stars \citep{2005ApJS..161..240T,2015ApJ...801L..28K}.
XMP galaxies are not primeval, but galaxies with an
underlying evolved stellar population 
\citep[e.g.,][]{2007ApJ...667L.151A,2008ApJ...675..194C,2010MNRAS.404.2037P,
2013AJ....146..144A} that look young because they  
may have recently accreted metal-poor gas 
\citep{2001A&A...374..800O,2010MNRAS.406.1238E,2013ApJ...767...74S,2014ApJ...783...45S,2015ApJ...810L..15S}, 
similar to the cold-flows predicted by the numerical simulations 
of galaxy formation 
\citep{2009Natur.457..451D,2010MNRAS.402.1536S,2014A&ARv..22...71S,2014MNRAS.441.2593N}.
Then the gas 
forming stars in XMPs would trace the cosmic web in their immediate 
surrounding, providing a method to study and to characterize 
this elusive component of the cosmic web.

Given the general interest and the potential of XMPs as astronomical 
tools, there have been many attempts to look for new candidates, i.e., to 
enlarge the list of XMPs, but also to find the least metallic object   in the local Universe. 
However, despite the unquestionable success in finding new objects, 
the club of  XMPs is still quite exclusive. The review 
paper by \citet{2000A&ARv..10....1K} contains only 31 targets,
\citet{2003ApJ...593L..73K} add 8 new targets from the early data 
release of SDSS, the search in SDSS-DR6 by
\citet{2009A&A...505...63G} yields 44, 
and the systematic 
bibliographic search for all XMPs in the literature carried out 
by
\citet[][hereinafter \paperi]{paperi} 
renders 140.
(\paperi\ includes targets found by
\citeauthor{2004ApJS..153..429K}~\citeyear{2004ApJS..153..429K},
\citeauthor{2004A&A...415...87I}~\citeyear{2004A&A...415...87I},
\citeauthor{2006A&A...448..955I}~\citeyear{2006A&A...448..955I},
and
\citeauthor{2007ApJ...665.1115I}~\citeyear{2007ApJ...665.1115I}.)
New local metal-poor objects have been discovered since the 
publication of \paperi\
\citep[e.g.,][]{2012A&A...546A.122I,2013AJ....146....3S,2015MNRAS.448.2687J,2015A&A...579A..11G},
but the situation has not changed in essence. 
Moreover, even though the number of members of the 
club has increased since the discovery 
of  IZw18 45 years ago 
\citep{1970ApJ...162L.155S,1972ApJ...173...25S},
its low metallicity has not been superseded; the present record-breaking object
SBS 0335-052W \citep{2009A&A...503...61I}
has a metallicity close to that of IZw18 and around 2\,\%\ of the 
solar value. The existence of such a lower limit for the metallicity of 
the XMPs has challenged interpretation for decades
\citep[e.g.,][]{2011EAS....48...95K}.

The low number of observed XMPs poses a problem since XMPs are
expected to be the most common galaxies in the local 
universe. Because of the luminosity-metallicity relationship 
\citep[e.g.,][]{1989ApJ...347..875S,2006ApJ...647..970L},  all galaxies fainter 
than absolute magnitude ${\rm M}_B\simeq -12.5$ are  expected to be XMPs
\citep[][]{2012ApJ...754...98B}. 
The extrapolation of the observed galaxy luminosity function to the faint end 
implies that these ultra-faint XMPs should be the most numerous galaxies in the 
local universe \citep[e.g.,][]{2005ApJ...631..208B,2015MNRAS.451.1540L}.
The reason why these objects are not detected in large quantities
seems to be an observational bias, associated with the fact that they 
are both faint and of low surface brightness. Only when they undergo  a 
starburst phase, brightness and surface density increase to make them
detectable in typical magnitude-limited optical surveys 
\citep[see][and references therein]{2013AJ....146....3S,2015MNRAS.448.2687J}.
In order to explain the supposed lack of XMPs, \citet[][]{2013AJ....146....3S} proposed 
the existence of two  classes of XMPs: 
(1)  the faint, low surface brightness, {\em quiescent} XMPs, 
very numerous but usually undetected,  
and (2) the {\em bursting} XMPs, going 
through a transit starburst  which allow them 
to appear in optical surveys like the Sloan Digital Sky Survey 
analyzed here \citep[][]{2002AJ....123..485S}. 
Unless otherwise explicitly stated, 
this paper refers to the latter, even though we often call them XMPs for simplicity.
We will mention the class of quiescent XMPs in connection with the 
number density estimate in Sect.~\ref{number_density}.

With a few exceptions, the observed
XMPs are bursting XMPs, 
that makes them blue and compact when the 
star-forming region is large with respect to the size of the galaxy
\citep[][]{2000A&ARv..10....1K}.
Thus bursting XMPs tend to be blue compact dwarf (BCD) galaxies.
Their optical spectrum is characterized by intense emission lines 
\citep[e.g.,][]{1991A&AS...91..285T}. Many of these XMPs
have cometary or tadpole morphology 
\citep[][\paperi]{2008A&A...491..113P}. 
XMP galaxies are gas rich, with
a typical gas-mass to stellar-mass ratio in excess of 10 
\citep{2013A&A...558A..18F}.
They tend to be in relative isolation, although some of them
show loose companions. They prefer 
low density environments identified as voids and sheets 
in cosmological numerical simulations \citep{2015ApJ...802...82F}.

One of the difficulties to enlarge the sample of XMPs is related to the determination
of their chemical abundances, which critically depends on the available optical emission 
lines.  The most accurate method relies on measuring electron temperatures,
and uses weak and near-UV emission lines often unavailable. 
In addition, tailored data-mining techniques are required to identify them 
since XMPs appear as rare objects in very large surveys.
Here we report a new systematic search for XMP galaxies in the 
Sloan Digital Sky Survey \citep[SDSS-DR7;][]{2009ApJS..182..543A}
that overcomes the two aforementioned difficulties.
The philosophy remains very much like in \paperi . We carry out an automated 
classification of the SDSS spectra using k-means
\citep[e.g.,][]{2010ApJ...714..487S,2014A&A...565A..53O} 
in a range of wavelengths 
particularly sensitive to metallicity. One or a few
classes have spectra characteristics of low-metallicity galaxies. 
Then we measure individually 
the metallicity of the objects in these classes, so as to confirm
or discard these XMP candidates  as true members of the XMP 
party. 
The difference between \paperi~ and the present work is in the 
wavelength region chosen to classify the spectra. In \paperi~ we used the region 
around H$\alpha$, where the two [NII] lines tend to 
disappear when the metallicity goes to zero
\citep[e.g.,][]{2004MNRAS.348L..59P,2012ApJ...756..163S}.
However, even objects with significant [NII] can be
XMP \citep{2014ApJ...797...81M}, 
because depending on the evolutionary state and
chemical history galaxies may have enhanced N/O, thus presenting 
large [NII] for a given O \citep[e.g.,][]{2009MNRAS.398..949P}.
Here we use the region between the lines [OIII]$\lambda$4363 and
[OIII]$\lambda$5007
which is known to depend  on the temperature of the 
HII region producing the emission-line spectrum. Since 
metals are a main coolant agent of the gas 
\citep[e.g.,][]{1979MNRAS.189...95P}, metal-poor objects 
are also high-temperature objects, yielding a relationship between 
electron temperature and metallicity.
This approach is similar to the one already used by
\citet{2015ApJ...805...45L}, except that they employ it to 
search for galaxies at a redshift around  0.8, so that the 
wavelength region 
of [OIII] is observed in the near IR.

The paper is organized as follows. First, we describe the
automated search for XMP candidates using k-means (Sect.~\ref{kmeanssearch}).
It leads to \ncandidates\ candidates. Then,  we use the SDSS spectra of the candidates 
to measure their integrated  metallicity, which narrows down the list  to \nxmps\ XMPs 
(Sect.~\ref{confirmation}). Emission line fluxes and abundances
are calculated in Sects.~\ref{fluxes} and \ref{hcm}, respectively. 
The final  list  with \nxmps\ XMPs is worked out in Sect.~\ref{list_of_xmp}
 (Table~\ref{main_list}),  where we also include a second list (Table~\ref{candidates}) 
with \nxmpsdos\ potential XMPs selected under less restrictive noise constraints. 
The global properties of the XMPs are analyzed in Sect.~\ref{properties},
explicitly,  stellar mass (Sect.~\ref{masses}),
reddening and  dust mass (Sect.~\ref{reddening}),  
absolute magnitude and color (Sect.~\ref{cmd}),
number density  (Sect.~\ref{number_density}),
metallicity threshold (Sect.~\ref{threshold}),
morphology (Sect.~\ref{morphology}),
underlying stellar population  (Sect.~\ref{agess}),
and large scale environment (Sect.~\ref{environment}).
The results are summarized and discussed in Sect.~\ref{discussion}.  
The number of false positives and false negatives in Table~\ref{main_list}
is estimated  in Appendix~\ref{appendix}.

%
\section{Search for XMP candidates classifying the spectrum around
[OIII]$\lambda$4363}\label{kmeanssearch}

The ratio between the emission of [OIII]$\lambda$4363 and the pair 
[OIII]$\lambda\lambda$4959,5007 strongly depends on the metallicity of the 
emitting gas. The reason is well known and rather straight-forward.  
Since metals are a main coolant of the gas \citep[e.g.,][]{1979MNRAS.189...95P}, 
metal-poor gas is also high-temperature gas. The ratio  [OIII]$\lambda$4363 to 
[OIII]$\lambda\lambda$4959,5007 is used to measure  [OIII] electron temperature in 
HII regions \citep[e.g.,][]{2008MNRAS.383..209H}, therefore it is also a good proxy 
for metallicity. The three [OIII] lines are excited by collisions with free electrons, 
and their decay through photon emission renders the observed lines. 
The excitation potential of the upper level of  [OIII]$\lambda$4363 ($^1{\rm S}_0$; $5.35$\,eV) 
is significantly higher than the upper level shared by [OIII]$\lambda$4959 
and [OIII]$\lambda$5007 ($^1{\rm D}_2$; $2.51$\,eV), so that [OIII]$\lambda$4363 increases 
relative to the other two lines as the temperature increases, that is to say, as the 
metallicity of the gas decreases. The relative fluxes change by more than
one order of magnitude when the electron temperature changes in the range 
of the typical temperatures in HII regions, say, 
between 5 and 25 kK \citep[e.g.,][]{1974agn..book.....O}.

\begin{figure}
\includegraphics[width=0.5\textwidth]{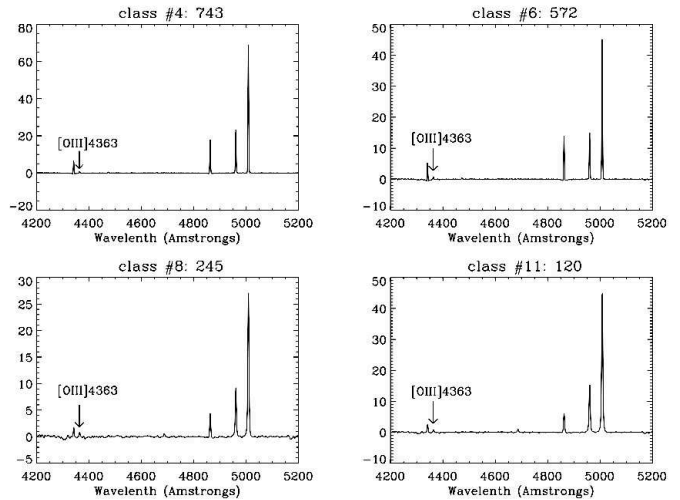}
\caption{Template of four spectral classes where the emission line 
[OIII]$\lambda$4363 (marked with arrows)  is particularly intense compared to   
[OIII]$\lambda$4959 and  [OIII]$\lambda$5007. They were found by applying 
two nested k-means classifications to 
the \ndrts\ galaxies with spectra in SDSS-DR7 having redshift $< 0.25$.
Two of the classes ( \#8 and \#11) are AGNs, as proven by their position 
on the BTP diagram and the fact that the emission lines are particularly
broad.
The spectra, with the continua removed, have been 
normalized to the peak intensity of  [OIII]$\lambda$4363. 
Wavelengths are given in \AA . The 
label on top of each panel represent the class number and 
the number of galaxies in the class.
}
\label{classes}
\end{figure}

We take advantage of this sensitivity of the spectrum on metallicity 
to search for XMP candidates. 
Using the automatic classification algorithm k-means
\citep[][]{2010ApJ...714..487S,2014A&A...565A..53O},
we classify the full SDSS-DR7 spectral database according to the
shape of the spectrum of the galaxies in the spectral region of
interest, from 4200 to 5200\,\AA , which contains 
[OIII]$\lambda$4363, [OIII]$\lambda$4959,  and [OIII]$\lambda$5007.
As in \paperi , the signal to be classified, $S(\lambda)$, is the observed spectrum 
in rest-frame wavelength, $I(\lambda)$, after removing the continuum intensity, 
$I_c(\lambda)$, and after the subsequent  normalization to the emission at the reference  
wavelength  (4363\,\AA\ in this case), explicitly,
\begin{equation}
S(\lambda)={{I(\lambda)-I_c(\lambda)}\over{I(4363\,{\rm \AA})-I_c(4363\,{\rm \AA})}}.
\end{equation}
We start from the full set of \ndrts\ galaxy spectra from SDSS-DR7 having
redshift smaller than 0.25 \citep{2010ApJ...714..487S}. The classification
renders 17 types or classes. The two of them showing the largest [OIII]$\lambda$4363, 
which include  24200 objects, are re-classified to obtain 12 subclasses. 
Some of these subclasses collect noisy spectrum, and those are discarded 
so that k-means acts as a noise filter. Four of the subclasses show particularly intense  
[OIII]$\lambda$4363, as depicted in Fig.~\ref{classes}. (Note that, even in these cases, 
the flux of  [OIII]$\lambda$4363 is much smaller than the fluxes of  [OIII]$\lambda$4959,  
[OIII]$\lambda$5007
-- 
the spectra in  Fig.~\ref{classes} have [OIII]$\lambda$5007 some fifty times 
larger than [OIII]$\lambda$4363.)  
Classes \#8 and \#11 in Fig.~\ref{classes} are discarded because 
the mean spectrum of the class appears in the region of 
Active Galactic Nuclei (AGNs)  of the BPT diagram 
\citep{1981PASP...93....5B,2012ApJ...756..163S}.  Therefore,  their high excitation is 
produced by hard AGN radiation, rather than by lack of efficient coolants  
in a stellar-radiation excited HII region. 
The AGN nature of the discarded classes is corroborated by the large
width of their emission lines (cf. AGN classes, \#8 and \#11, with 
star-forming classes  \#4 and \#6), a property also 
characteristic of AGNs. After the classification and trimming, we are
left with \ncandidates\ 
bursting XMP candidates.  

We have compared these \ncandidates\ candidates with the 32~XMPs
selected in \paperi\ after classifying the spectral region around
H$\alpha$. Most of them, 29  out of the 32, are included in our list of candidates. 
Considering that the two procedures are completely independent,
and the randomness inherent to the k-means procedure, 
the agreement is very reasonable. Recall that k-means is a powerful tool able to classify
very large data sets in high-dimensional spaces. However,  among
its drawbacks, k-means does not provide a single classification for
a single database. Each particular run depends on a random 
initialization, thus, the classification is not unique 
\citep[e.g.,][Sect.~2]{2010ApJ...714..487S}.

Nothing ensures that the \ncandidates\ candidates are true XMPs. 
They have been picked up as objects where  [OIII]$\lambda$4363 is 
particularly intense, using as reference the rest of galaxies in the 
spectroscopic database of SDSS-DR7.
In order to show whether or not they are XMPs, we have 
computed the metallicity for every one of them 
to mark as XMP only those objects of truly low metallicity.
(The details are  given in the next Sect.~\ref{confirmation}.)
One may think of this second step as a way to clean 
from pollutants the systematic but otherwise rough XMP selection carried out by 
k-means on the SDSS-DR7 database.

%
\section{Metallicities of the XMP candidates}\label{confirmation}

\subsection{Determination of emission line fluxes}\label{fluxes}

We use the SDSS-DR7 spectra 
of the \ncandidates\ candidates to 
determine their metallicities. As usual, the procedure requires correcting the emission 
line fluxes for reddening, which is determined from the relative fluxes of the observed Balmer lines
\citep[e.g.,][]{2008MNRAS.383..209H,2012EAS....54....3S}. However,
Balmer emission is usually depressed due to the presence of absorption lines of stellar origin. 
Therefore, prior to the measurement of the emission lines, we perform a stellar fitting to the 
continuum of the spectra. We use {\sc starlight} \citep{2005MNRAS.358..363C}
to model the spectral energy distribution (SED) of the underlying stellar population.
{\sc Starlight} fits the observed continuum as a linear combination of single stellar 
population  (SSP)
spectra of various ages and metallicities.  We chose the SSP spectra 
from \citet{2003MNRAS.344.1000B}, including four metallicities 
$Z = 0.0001, 0.0004, 0.004,$ and 0.008
\citep[1\%, 3\%, 30\% and 60\% $Z_\odot$, respectively;][]{2009ARA&A..47..481A}.
Reddening of the stellar SED was included assuming the extinction to follow  
\citet{1989ApJ...345..245C} with $R_{V}$ = 3.1.
The stellar continuum thus obtained is subtracted out of the observed 
spectrum to produce a pure emission line spectrum. The latter is employed to determine
line fluxes.

We used the package
{\sc shifu}\footnote{SHerpa IFU line fitting package, \citet{garciabenito+15}.} 
to obtain the flux of the emission 
lines. The package contains a suite of routines to easily analyze emission or absorption lines 
in IFU data (both in cube and RSS format). Individual spectra, such as those 
employed in our particular case, can be provided 
as a list of files (RSS). The core of the code uses CIAO's Sherpa package 
\citep{2001SPIE.4477...76F,2007ASPC..376..543D}. Several custom algorithms are implemented in order
to cope with general and ill-defined cases. A sigma-clipping is independently applied to the 
stellar continuum model, and then this is parsed to the composite line plus continuum model. 
Although the fitting is 
performed in the residual spectra, we allow for the modeling of the continuum to take into account small 
deviations in the stellar continnum residuals. A first order polynomial has been chosen for the continuum, 
while single gaussians have been selected for the lines. 
The continuum is evaluated in the  original spectra when used
to determine equivalent widths.

\subsection{Method to measure metal abundance}\label{hcm}

Metal abundances are computed with 
the code HII-CHI-mistry\footnote{{\tt http://www.iaa.es/$\sim$epm/HII-CHI-mistry.html}}
  \citep[hereafter HCM][]{2014MNRAS.441.2663P},
which calculates the total oxygen abundance O/H, the nitrogen-to-oxygen ratio N/O,
and the ionization parameter U. 
Normalized to H$\beta$ and corrected for extinction, the  fluxes of the emission lines
[OII]$\lambda$3727, [OIII]$\lambda\lambda$4363,5007, [NII]$\lambda$6584, and
[SII]$\lambda\lambda$6716,6731 are compared with the predictions of a large grid of 
{\sc cloudy} \citep{2013RMxAA..49..137F} photoionization models.
The HCM assumes a spherically symmetric constant-density distribution 
of gas ionized by {\sc popstar} \citep{2009MNRAS.398..451M}
1\,Myr-old stellar clusters with the same metallicity of the gas, 
and covering a wide range of values in O/H, N/O, and log U.
In particular, the grid includes very low values of the gas metallicity,
down to 12+log(O/H)=6.9 equivalent to 1/60 of the oxygen 
abundance in the solar composition \citep{2009ARA&A..47..481A}.   
The main advantage of this method is that it leads to a derivation
of the chemical abundances consistent with the direct
method even in the absence of one or more of the input emission lines
\citep{2014MNRAS.441.2663P}.
Since the direct method does not depend on modeling 
\citep[e.g.,][]{1974agn..book.....O,2012EAS....54....3S}, it means that the 
metallicities provided by HCM are almost insensitive to the 
model assumptions.
Given a set of observed line ratios, HCM assigns a $\chi^2$ to each 
one of the models, computed as the sum of 
the square differences between the observed and 
the model line ratios.  The $1/\chi^2$-weighted 
average of the abundance in the models
yields the chemical abundance tagged to  the observed line ratios.
Errors are derived as the $1/\chi^2$-weighted standard 
deviation of the parameters in the models. When the 
temperature-sensitive line [OIII]$\lambda$4363 is available, 
HCM uses the whole grid of models to provide values of O/H and 
N/O  that are consistent with the direct method, even in the 
absence of [OII]$\lambda$3727. Thus the use of HCM results 
especially appropriate to search for XMPs in SDSS since 
[OII]$\lambda$3727 is sometimes outside the observed 
spectral region for nearby objects.  
The error bars estimated by HCM quantify the internal precision of the 
method, i.e., they give the range of models that is consistent with the 
observed line fluxes. The observational errors enter into 
the abundance error estimate indirectly, since the observed 
fluxes must have enough signal to be considered for analysis.

\subsection{List of XMPs}\label{list_of_xmp}

\begin{figure*}
\includegraphics[width=\textwidth]{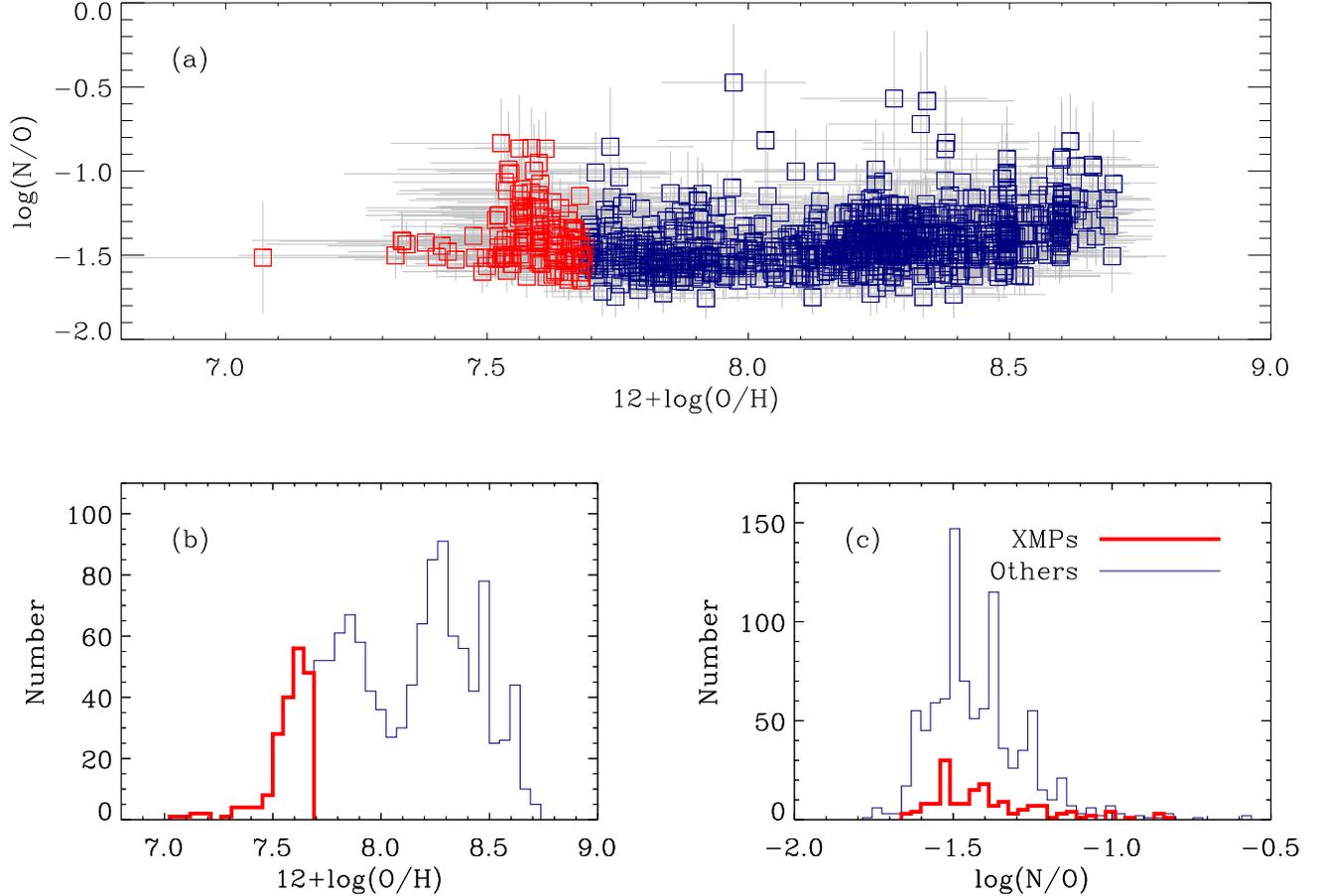}
\caption{
(a) Scatter plot N/O vs O/H  for the XMPs found in our search 
(red squares with error bars), and for the candidates that turned out not to be XMPs 
(blue symbols with error bars).
(b) Distribution of the O abundance for the XMPs (thick red line) and
the failed candidates (thin blue line). It is rather smooth at the transition
abundance (7.69, equivalent to a tenth of the solar value). 
(c) Distribution of N/O for the XMPs (thick red line) and the rest (thin blue line).
}
\label{distribution}
\end{figure*}

The automated procedure described in Sect.~\ref{fluxes} provides the fluxes used by 
HCM (Sect.~\ref{hcm}) to infer the oxygen abundance of the \ncandidates\ candidates
 (Sect.~\ref{kmeanssearch}). 
We consider only line fluxes with a signal-to-noise ratio larger than three,
including the weak [OIII]$\lambda$4363. The scatter plot 12+log(O/H) vs 
log(N/O) for the full sample is shown in Fig.~\ref{distribution}a. Assuming that the solar composition 
corresponds to 12+log(O/H)$_\odot\simeq 8.69$~\citep{2009ARA&A..47..481A}, 
XMPs are defined as those candidates having\footnote{Different authors 
use slightly different thresholds, e.g., \citet{2000A&ARv..10....1K} and 
\citet{2011ApJ...743...77M} use 
$12+\log({\rm O/H}) \le\ 7.65$, \citet{2009A&A...505...63G} use
$\leq 7.60$,  while we use exactly one tenth of the solar metallicity quoted in the 
review paper by \citet{2009ARA&A..47..481A}.\label{thresholds_note}}
\begin{equation}
12+\log({\rm O/H})\le 7.69. 
\label{solar_metal}
\end{equation}
There are \nxmps\ galaxies fulfilling this requirement, 
which are the
bursting XMP galaxies 
presented in this work. 
\duplicated\ of them have duplicated spectra in SDSS-DR7. 
We have chosen for analysis the spectrum yielding the smallest error in  
metallicity. The XMP galaxies
are represented as red squares with 
error bars in Fig.~\ref{distribution}a. The distribution of metallicities of the candidates is
given in Fig.~\ref{distribution}b. The histogram is fairly continuous at the border between 
XMPs and the rest, which proves the 1/10 solar metallicity threshold to be rather arbitrary.
The list of XMPs is given in Table~\ref{main_list}, which includes coordinates 
as well as the main physical properties provided by HCM. 
When [NII]$\lambda$6584 is too weak for the noise level, 
N/O cannot be derived,  and it appears as -9.99 in the Table. 
The distributions of N/O for the XMPs and the rest of the sample 
are given in Fig.~\ref{distribution}c. They look qualitatively similar.

We carried out several sanity checks. First, we estimated the number of false positives 
and false negatives due to observational errors. False positives are galaxies in the 
XMP list which have true oxygen abundance above the threshold. False negatives are
true XMP galaxies that have been discarded from the selection. Using a Bayesian approach, 
we work out in Appendix~\ref{appendix} the expected number of false positives and 
false negatives. Even though the typical error of the individual abundance estimates 
is rather small ($\simeq 0.17$\,dex for the galaxies in Table~\ref{main_list}), 
we expect a significant number of false positives and false negatives. Assuming the
true distribution of metallicities to be given by the histogram in 
Fig.~\ref{distribution}b, we expect 82 false positives and 46 false negatives
(Eq.~[\ref{false3}]). If the true distribution is uniform, then 
the number of both false positives and false negatives is around 55 
(Eq.~[\ref{false4}]).  
Partly to offer the possibility of recovering false negatives in the future, 
we include a 2nd list with {\em candidate} XMPs  in Table~\ref{candidates}.
Table~\ref{main_list} is based on lines with measured fluxes exceeding 
three times the noise. If this threshold is lowered, and lines with 
fluxes at two-sigma are also considered, then the number of galaxies 
fulfilling the condition in Eq.~(\ref{solar_metal})
increases to  \nxmpsdos , and those are the objects included in 
Table~\ref{candidates}. The reason why the number increases when lowering 
noise limit can be pinned down to the weakness of [OIII]$\lambda$4363,
which tends to be excluded from the metallicity estimate when
the noise limit is too restrictive. When it is considered in the metallicity 
estimate,  HCM tends to lower the oxygen abundance estimate due to 
arguments given in  Sect.~\ref{kmeanssearch}

\begin{figure}
\includegraphics[width=0.45\textwidth]{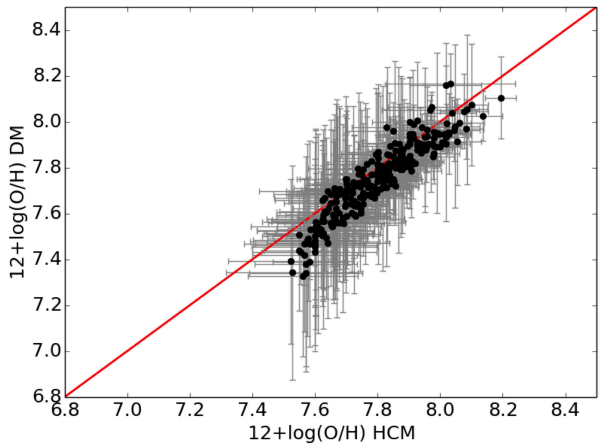}
\caption{Scatter plot direct-method metallicity vs HCM metallicity for 
the galaxies having [OII]$\lambda$3727 within the observed spectral 
range. Both measurements agree quite well, with a small difference at 
metallicities below the XMP threshold which consequently 
does not affect  the selection of XMPs.
The solid line  corresponds to the  one-to-one relationship.
}
\label{metal_vs_metal}
\end{figure}

As a second test, we compared the metallicity provided by HCM 
with the metallicity obtained with the direct method \citep[e.g.,][]{2008MNRAS.383..209H} 
when [OII]$\lambda$3727 was available
\footnote{The wavelength 3727\,\AA\ is out of the range 
of the original SDSS spectrograph, so that a minimum redshift
around 0.03 is needed for line to appear in a SDSS spectrum.
}. 
These abundances, based on the direct method, are also included in Table~\ref{main_list}. 
The scatter plot direct-method metallicity vs HCM metallicity is shown in 
Fig.~\ref{metal_vs_metal}. As expected, the two techniques agree quite well
with a dispersion of only 0.04\,dex, which is smaller than the intrinsic
error of both methods. On top of this agreement, there is a tendency for  
HCM to overestimate the direct-method metallicity at low metallicities. 
Since this bias occurs below the threshold metallicity to  
be XMP (Eq.~[\ref{solar_metal}]), 
both  methods would yield the same list of XMPs if we had access to 
[OII]$\lambda$3727.  
Moreover, the systematic differences between the abundances
obtained from HCM and the direct method are small compared
with their error bars,  so that such differences barely 
affect other results and conclusions of the paper.
For example, adding an extra 0.06~dex error to the 
typical $\sim 0.17~$dex error bar increases the
number of false positives worked out in Appendix~\ref{appendix}, but
only from 82 to 86. 

\begin{figure}
\includegraphics[width=0.45\textwidth]{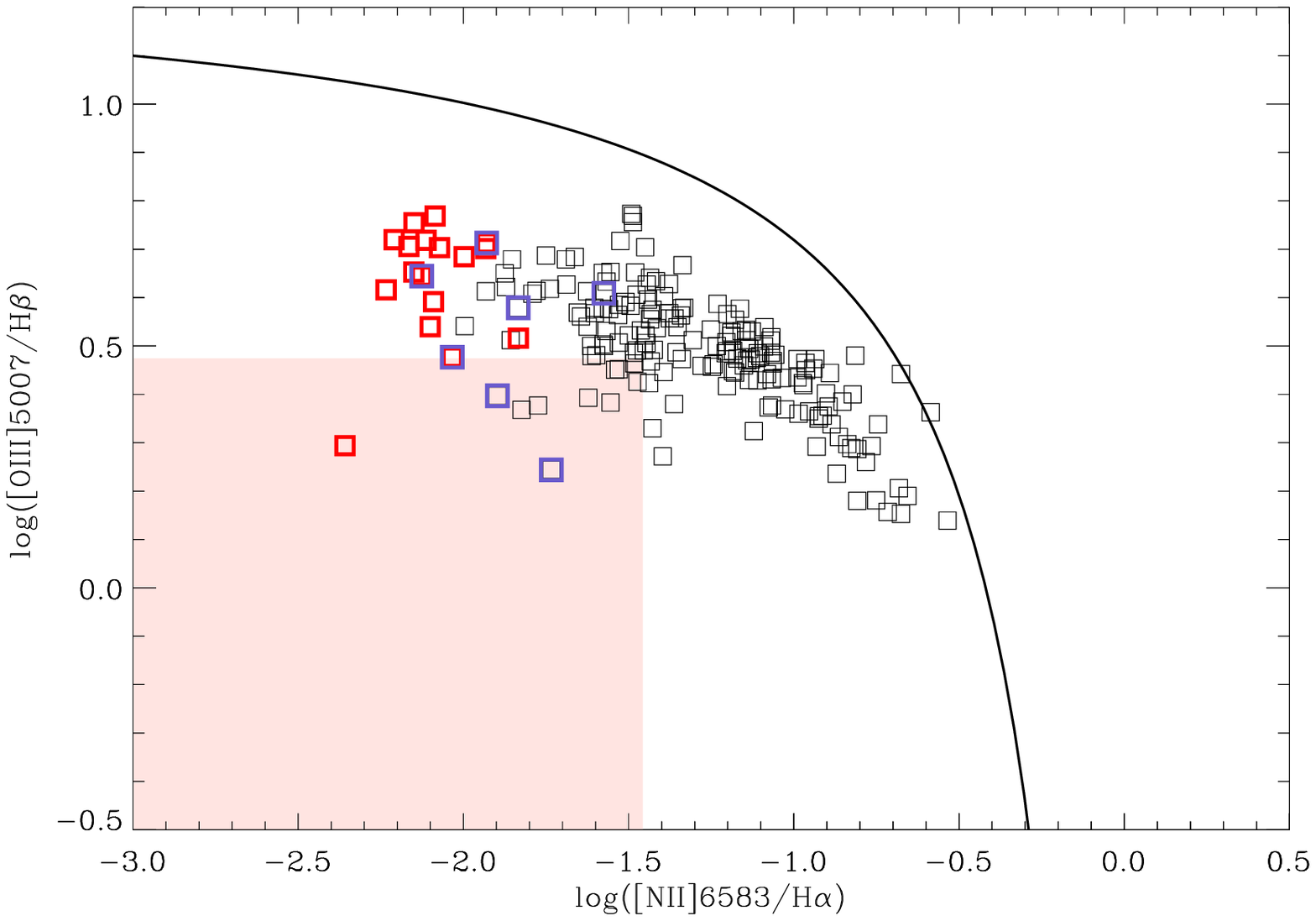}
\caption{
Scatter plot [OIII]$\lambda$5007/H$\beta$ vs 
[NII]$\lambda$6583/H$\alpha$ for the galaxies in Table~\ref{main_list}.
This plane is commonly used to separate starbursts from 
AGNs  \citep[the so-called BPT diagram, after][]{1981PASP...93....5B}. Most
objects  reside below the solid line by \citet{2003MNRAS.346.1055K}, 
corresponding to excitation  by stars.
Only two objects are close to this border and they may have composite spectra.
The shaded region shows the area used by \citet{2012A&A...546A.122I} 
to search for XMPs in SDSS-DR7. The colored points correspond to  
objects also found in \paperi\ (red) and in \citet[][blue]{2012A&A...546A.122I}.
Contrarily to our targets, these previous searches pick up emitters 
concentrated on the  leftmost region of the BPT diagram, 
corresponding to [NII]$\lambda$6583$\ll $H$\alpha$.
}
\label{bpt_diagram}
\end{figure}

As a final test, we represented our targets in the [OIII]$\lambda$5007/H$\beta$ vs 
[NII]$\lambda$6583/H$\alpha$ plane commonly used to separate starbursts from AGNs 
\citep[the so-called BPT diagram, after][]{1981PASP...93....5B}; see Fig.~\ref{bpt_diagram}. 
None of them are AGNs, and only two  reside in the composite-spectrum region 
next to the divide worked out by
\citet{2003MNRAS.346.1055K}.
They may be starbursts of high N/O \citep{2009MNRAS.398..949P}, or HII regions
where the excitation may have contribution from shocks, evolved stars,  
or even AGNs. The two objects with composite-spectrum 
are number 51 and number 151 in Table~\ref{main_list}, and their
identification as XMP galaxies must be taken with caution. 

%
Most of the \nxmps\ galaxies in Table~\ref{main_list} seem to be new 
members of the XMP class. Table~2 of \paperi\ compiles
all XMPs in the literature up to the date of publication in mid 2011. 
It contains 129 objects, 21 of which coincide with objects in Table~\ref{main_list}. 
\paperi\ also carries out a systematic search for XMPs in SDSS-DR7, 
finding 32 such objects. 17 of these 32 targets coincide with targets 
selected in our study, and so are included in Table~\ref{main_list}.  In addition,
there is a  recent major search for XMPs by \citet{2012A&A...546A.122I}, 
which contains 41 targets.  Among them, 7 are in Table~\ref{main_list}. 
Considering all the objects in these three lists, and discarding coincidences,  
only 31 among the \nxmps\ galaxies in Table~\ref{main_list} seem to be 
known.  Known and unknown targets are pointed out in one of the columns of
Table~\ref{main_list}, which used the following code to indicate matches:  
(1) Table~1 in \paperi , (2) Table 2 in \paperi , (3) Tables 1 \&\ 2 in \paperi , 
(4) \citet{2012A&A...546A.122I}, (5) Table 1  in \paperi\  \&\ \citet{2012A&A...546A.122I},
(6) Table 2  in \paperi\  \&\ \citet{2012A&A...546A.122I},
(7) Table 1 \&\ 2 in \paperi , \&\ \citet{2012A&A...546A.122I}, and finally,  
(0) no match.

Part of the difference with previous searches is related to the abundance 
threshold (Eq.~[\ref{solar_metal}]). Our criterion is slightly 
less restrictive than others (see footnote number~\ref{thresholds_note}) therefore,
by construction, the highest metallicity targets in Table~\ref{main_list}  are
discarded in other compilations. For example, 
43 and 88 objects have $12+\log({\rm O/H}) > 7.65$ and $> 7.60$, respectively. 
In addition to this fact, however, we employ a novel selection procedure 
which influences the results.    
Figure~\ref{bpt_diagram}  provides the clue for understanding the main 
difference between our search and previous ones. 
Previous searches are focused on objects having small 
[NII]$\lambda$6583 compared with H$\alpha$ (\paperi ) 
or with H$\beta$ \citep[e.g.,][]{2007ApJ...665.1115I,2012A&A...546A.122I},
whereas our technique selects objects that cover a significant part
of the BPT plane. XMP galaxies of relatively large [NII]$\lambda$6583/H$\alpha$
are know to exist due to, e.g., having particularly large N/O or displaying
particularly low ionization  \citep[e.g.,][]{2014ApJ...797...81M}. 
Our approach is sensitive to those XMPs, as it is  evidenced in Fig.~\ref{bpt_diagram}, 
which shows with colored symbols the objects in common with the searches in \paperi\ and 
\citet{2012A&A...546A.122I}.
The region explored by \citep{2012A&A...546A.122I} is also indicated in the 
figure.\footnote{The constraints in \citet{2012A&A...546A.122I} are  
[OIII]$\lambda$4959/H$\beta\lesssim 1$ and 
[NII]$\lambda$6583/H$\beta\lesssim 0.1$. 
However, atomic properties set  [OIII]$\lambda$5007/[OIII]$\lambda$4959$\simeq 2.98$ 
\citep[e.g.,][]{2000MNRAS.312..813S} and H$\alpha$/H$\beta\simeq 2.86$ 
\citep[e.g.,][case B recombination]{1989agna.book.....O}, 
which brings the constraints in the line ratios used for our BPT diagram to 
[OIII]$\lambda$5007/H$\beta\lesssim 2.98$ and 
[NII]$\lambda$6583/H$\alpha\lesssim 0.035$. This is the region highlighted in Fig.~\ref{bpt_diagram}.
}

The histogram of metallicities in Fig.~\ref{distribution}a drops at  
$12+\log({\rm O/H}) \le 7.50$, so that only 28 of the 
XMPs in Table~\ref{main_list} have oxygen abundance smaller than this 
threshold. (The lowest metallicity limit is discussed in Sect.~\ref{threshold}.)
Those objects  are particularly rare, here and in previous works 
\citep[e.g., \paperi ;][]{2012A&A...546A.122I,2015A&A...579A..11G},
however, this part of Table~\ref{main_list}
is the one that contains more targets in common with previous 
searches. Actually, 36\,\%\ of our XMPs with $12+\log({\rm O/H}) \le 7.50$ 
are already known, whereas the 31 known objects in 
Table~\ref{main_list} represent only 16\% of the full list of 
\nxmps\ targets.

%
\begin{deluxetable*}{cccccccc}
\tabletypesize{\scriptsize}
\tablewidth{0pt}
\tablecaption{Extremely metal poor galaxies selected in this study\tablenotemark{a}}
\tablehead{
\colhead{Index\tablenotemark{b}}&
\colhead{Name\tablenotemark{c}}&
\colhead{12+log(O/H)\tablenotemark{d}}&
\colhead{log(N/O)\tablenotemark{d}}&
\colhead{logU\tablenotemark{e}}&
\colhead{CH$\beta$\tablenotemark{f}}&
\colhead{Match\tablenotemark{g}}&
\colhead{12+log(O/H)$_{\rm DM}$\tablenotemark{h}}
}
\startdata
  1&J$001637.4+003828.8$&7.66$\pm$0.18&-1.60$\pm$0.14&-2.65$\pm$0.12&0.00$\pm$0.04&0&7.59$\pm$0.19\\
  2&J$001922.9-002450.5$&7.57$\pm$0.19&-1.11$\pm$0.23&-2.54$\pm$0.16&0.27$\pm$0.03&0&7.47$\pm$0.18\\
  3&J$002114.8-085809.5$&7.56$\pm$0.23&-1.14$\pm$0.28&-2.65$\pm$0.19&0.62$\pm$0.01&0&9.99$\pm$0.00\\
  4&J$002916.8-010021.2$&7.61$\pm$0.16&-1.35$\pm$0.15&-2.04$\pm$0.33&0.17$\pm$0.02&0&9.99$\pm$0.00\\
  5&J$003630.4+005234.7$&7.58$\pm$0.12&-9.99$\pm$0.00&-1.88$\pm$0.25&0.01$\pm$0.03&1&7.39$\pm$0.08\\
  6&J$003741.1+003320.1$&7.68$\pm$0.12&-1.52$\pm$0.13&-1.89$\pm$0.27&0.00$\pm$0.03&0&9.99$\pm$0.00\\
  7&J$004224.8+003315.7$&7.63$\pm$0.16&-1.51$\pm$0.14&-2.55$\pm$0.18&0.07$\pm$0.03&0&9.99$\pm$0.00\\
  8&J$005249.8-084133.9$&7.63$\pm$0.15&-1.48$\pm$0.15&-2.66$\pm$0.21&0.23$\pm$0.03&0&9.99$\pm$0.00\\
  9&J$011914.3-093546.2$&7.60$\pm$0.14&-1.55$\pm$0.12&-1.92$\pm$0.27&0.15$\pm$0.02&2&9.99$\pm$0.00\\
 10&J$013352.6+134209.4$&7.46$\pm$0.14&-9.99$\pm$0.00&-1.79$\pm$0.23&0.26$\pm$0.03&2&9.99$\pm$0.00\\
 11&J$015759.7+011138.6$&7.56$\pm$0.29&-1.22$\pm$0.20&-2.65$\pm$0.13&0.11$\pm$0.05&0&9.99$\pm$0.00\\
 12&J$022107.2-002507.3$&7.60$\pm$0.20&-1.41$\pm$0.16&-2.53$\pm$0.19&0.15$\pm$0.02&0&9.99$\pm$0.00\\
 13&J$022237.8+002908.0$&7.67$\pm$0.15&-1.56$\pm$0.13&-2.54$\pm$0.17&0.28$\pm$0.05&0&9.99$\pm$0.00\\
 14&J$022607.0+000017.7$&7.64$\pm$0.20&-1.39$\pm$0.17&-2.56$\pm$0.17&0.04$\pm$0.02&0&9.99$\pm$0.00\\
 15&J$024223.6-005817.3$&7.66$\pm$0.12&-1.29$\pm$0.16&-1.78$\pm$0.19&0.21$\pm$0.02&0&7.54$\pm$0.17\\
 16&J$024429.0+000741.4$&7.55$\pm$0.12&-1.01$\pm$0.00&-1.64$\pm$0.18&0.00$\pm$0.06&0&9.99$\pm$0.00\\
 17&J$024636.4+001654.9$&7.60$\pm$0.20&-1.42$\pm$0.16&-2.53$\pm$0.19&0.30$\pm$1.27&0&9.99$\pm$0.00\\
 18&J$025818.9+004348.7$&7.60$\pm$0.24&-1.41$\pm$0.18&-2.53$\pm$0.18&0.26$\pm$0.01&0&9.99$\pm$0.00\\
 19&J$030149.0-005257.4$&7.58$\pm$0.12&-9.99$\pm$0.00&-1.86$\pm$0.24&0.29$\pm$0.04&2&9.99$\pm$0.00\\
 20&J$031137.9-003332.0$&7.58$\pm$0.13&-9.99$\pm$0.00&-2.38$\pm$0.20&0.19$\pm$0.01&0&9.99$\pm$0.00\\
\vdots&\vdots&\vdots&\vdots&\vdots&\vdots&\vdots&\vdots
\enddata
\tablenotetext{a}{The full table with all the \nxmps\ entries is given only in electronic format.}
\tablenotetext{b}{Sorted according to growing RA.}
\tablenotetext{c}{RA and DEC in J2000 coordinates.}
\tablenotetext{d}{Using HCM; see Sect.~\ref{hcm}.}
\tablenotetext{e}{Ionization parameter \citep[e.g.,][]{2012EAS....54....3S}.}
\tablenotetext{f}{Extinction coefficient in H$\beta$.}
\tablenotetext{g}{Galaxy also in: (1) Table 1 in \paperi ; (2) Table 2 in \paperi ; (3) Tables 1 \&\ 2 in \paperi ; 
(4) \citet{2012A&A...546A.122I}; (5) Table 1  in \paperi\  \&\ \citet{2012A&A...546A.122I};
(6) Table 2  in \paperi\  \&\ \citet{2012A&A...546A.122I};
(7) Table 1 \&\ 2 in \paperi , \&\ \citet{2012A&A...546A.122I}; (0) no match.}
\tablenotetext{h}{Using the direct method (Sect.~\ref{list_of_xmp}). 9.99 means no measurement available}
\label{main_list}
\end{deluxetable*}

%
\begin{deluxetable*}{cccccc}
\tabletypesize{\scriptsize}
\tablewidth{0pt}
\tablecaption{Candidate XMP galaxies\tablenotemark{a}}
\tablehead{
\colhead{Index\tablenotemark{b}}&
\colhead{Name\tablenotemark{c}}&
\colhead{12+log(O/H)\tablenotemark{d}}&
\colhead{log(N/O)\tablenotemark{d}}&
\colhead{logU\tablenotemark{e}}&
\colhead{CH$\beta$\tablenotemark{f}}
}
\startdata
 1&J$000009.0+011423.8$&7.67$\pm$0.18&-1.44$\pm$0.15&-2.58$\pm$0.13&0.29$\pm$0.06\\
  2&J$001432.7+002919.9$&7.60$\pm$0.24&-1.37$\pm$0.22&-2.63$\pm$0.12&0.09$\pm$0.03\\
  3&J$001637.4+003828.8$&7.66$\pm$0.18&-1.60$\pm$0.14&-2.65$\pm$0.12&0.00$\pm$0.04\\
  4&J$001922.9-002450.5$&7.57$\pm$0.19&-1.11$\pm$0.23&-2.54$\pm$0.16&0.27$\pm$0.03\\
  5&J$002114.8-085809.5$&7.56$\pm$0.23&-1.14$\pm$0.28&-2.65$\pm$0.19&0.62$\pm$0.01\\
  6&J$002535.1+003456.1$&7.61$\pm$0.18&-1.27$\pm$0.24&-2.68$\pm$0.21&0.15$\pm$0.04\\
  7&J$002916.8-010021.2$&7.61$\pm$0.16&-1.35$\pm$0.15&-2.04$\pm$0.33&0.17$\pm$0.02\\
  8&J$003145.3-110656.8$&7.61$\pm$0.17&-1.56$\pm$0.12&-2.24$\pm$0.33&0.13$\pm$0.18\\
  9&J$003630.4+005234.7$&7.59$\pm$0.12&-1.51$\pm$0.14&-1.89$\pm$0.25&0.01$\pm$0.03\\
 10&J$003741.1+003320.1$&7.68$\pm$0.12&-1.52$\pm$0.13&-1.89$\pm$0.27&0.00$\pm$0.03\\
 11&J$004224.8+003315.7$&7.63$\pm$0.16&-1.51$\pm$0.14&-2.55$\pm$0.18&0.07$\pm$0.03\\
 12&J$004521.9-093700.1$&7.52$\pm$0.21&-1.44$\pm$0.13&-1.92$\pm$0.43&0.09$\pm$0.03\\
 13&J$004614.4+000635.9$&7.55$\pm$0.20&-1.28$\pm$0.25&-2.50$\pm$0.19&0.25$\pm$0.03\\
 14&J$005249.8-084133.9$&7.63$\pm$0.15&-1.48$\pm$0.15&-2.66$\pm$0.21&0.23$\pm$0.03\\
 15&J$010408.4-004010.3$&7.60$\pm$0.15&-1.38$\pm$0.16&-2.38$\pm$0.17&0.21$\pm$0.12\\
 16&J$010413.3+001529.0$&7.56$\pm$0.21&-1.20$\pm$0.22&-2.65$\pm$0.13&0.16$\pm$0.04\\
 17&J$010414.6-005040.1$&7.31$\pm$0.34&-1.38$\pm$0.22&-1.71$\pm$0.32&0.20$\pm$0.04\\
 18&J$011144.5+010929.8$&7.55$\pm$0.22&-1.44$\pm$0.23&-2.41$\pm$0.24&0.09$\pm$0.23\\
 19&J$011340.5+005239.2$&7.04$\pm$0.26&-9.99$\pm$0.00&-1.55$\pm$0.10&0.00$\pm$0.01\\
 20&J$011914.3-093546.2$&7.60$\pm$0.14&-1.55$\pm$0.12&-1.92$\pm$0.27&0.15$\pm$0.02\\
\vdots&\vdots&\vdots&\vdots&\vdots&\vdots
\enddata
\tablenotetext{a}{The full table with the \nxmpsdos\ candidates is given
only in electronic format.}
\tablenotetext{b}{Sorted according to growing RA.}
\tablenotetext{c}{RA and DEC in J2000 coordinates.}
\tablenotetext{d}{Using HCM; see Sect.~\ref{hcm}.}
\tablenotetext{e}{Ionization parameter \citep[e.g.,][]{2012EAS....54....3S}.}
\tablenotetext{f}{Extinction coefficient in H$\beta$.}
\label{candidates}
\end{deluxetable*}

%
\section{Properties of the XMPs }\label{properties}

%
\subsection{Reddening and dust mass}\label{reddening}

Dust is made out of metals and so metal-poor objects are also 
expected to be dust-poor 
\citep[][]{2013MNRAS.436.2747K,2014Natur.505..186F}. 
However, 
the type of  bursting XMPs selected here are going through 
a starburst phase where SNe go off continuously creating 
dust that may potentially obscure the light from the 
starburst
\citep[e.g.,][]{2013ApJ...778..159T}.
Consequently, one expects small but non-zero amounts of extinction in 
XMPs. This is actually the case in our XMP sample.

As a by-product of the metallicity estimate, we have to measure the
extinction coefficient that renders the observed Balmer decrement (Sect.~\ref{fluxes}).
We follow the procedure described by \citet{2008MNRAS.383..209H},
where the shape of the extinction law is taken to be the Milky-Way 
extinction law by \citet{1989ApJ...345..245C}. 
Using the usual parametrization, the extinction 
coefficient at H$\beta$, c(H$\beta$), is defined
to be 
\begin{equation}
{\rm c(H}\beta)=-\log(F/F_0),
\end{equation}
where the symbols $F$ and $F_0$ stand for the reddened and original
fluxes in H$\beta$, respectively.  Therefore, the extinction coefficient 
is proportional to the  optical depth along the line-of-sight $\varpi$
\citep[e.g.,][]{1974agn..book.....O},
\begin{displaymath}
{\rm c(H\beta)}=\varpi\,\log e,
\end{displaymath}
with 
\begin{equation}
\varpi=\int_0^{l_0}\kappa_D\,\rho_D\,dl.
\label{def_tau}
\end{equation}
The integral goes along the line-of-sight from the observer 
to the source of light at $l_0$. The symbols  $\rho_D$ and 
$\kappa_D$  stand for the dust density and the cross 
section per unit mass of  dust, respectively.
Defining the density-weighted mean cross section $\langle\kappa_D\rangle$, 
\begin{equation}
\langle\kappa_D\rangle=\int_0^{l_0}\kappa_D\,\rho_D\,dl\Big/\int_0^{l_0}\rho_D\,dl,
\end{equation}
then the dust-mass column density, 
\begin{displaymath}
\Sigma_D=\int_0^{l_0}\rho_D\,dl,
\end{displaymath}
can be inferred from the extinction coefficient as 
\begin{equation}
\Sigma_D={\rm c(H\beta)}/\big[\langle\kappa_D\rangle\log e\big].
\label{dust_mass}
\end{equation}

\begin{figure}
\includegraphics[width=0.5\textwidth]{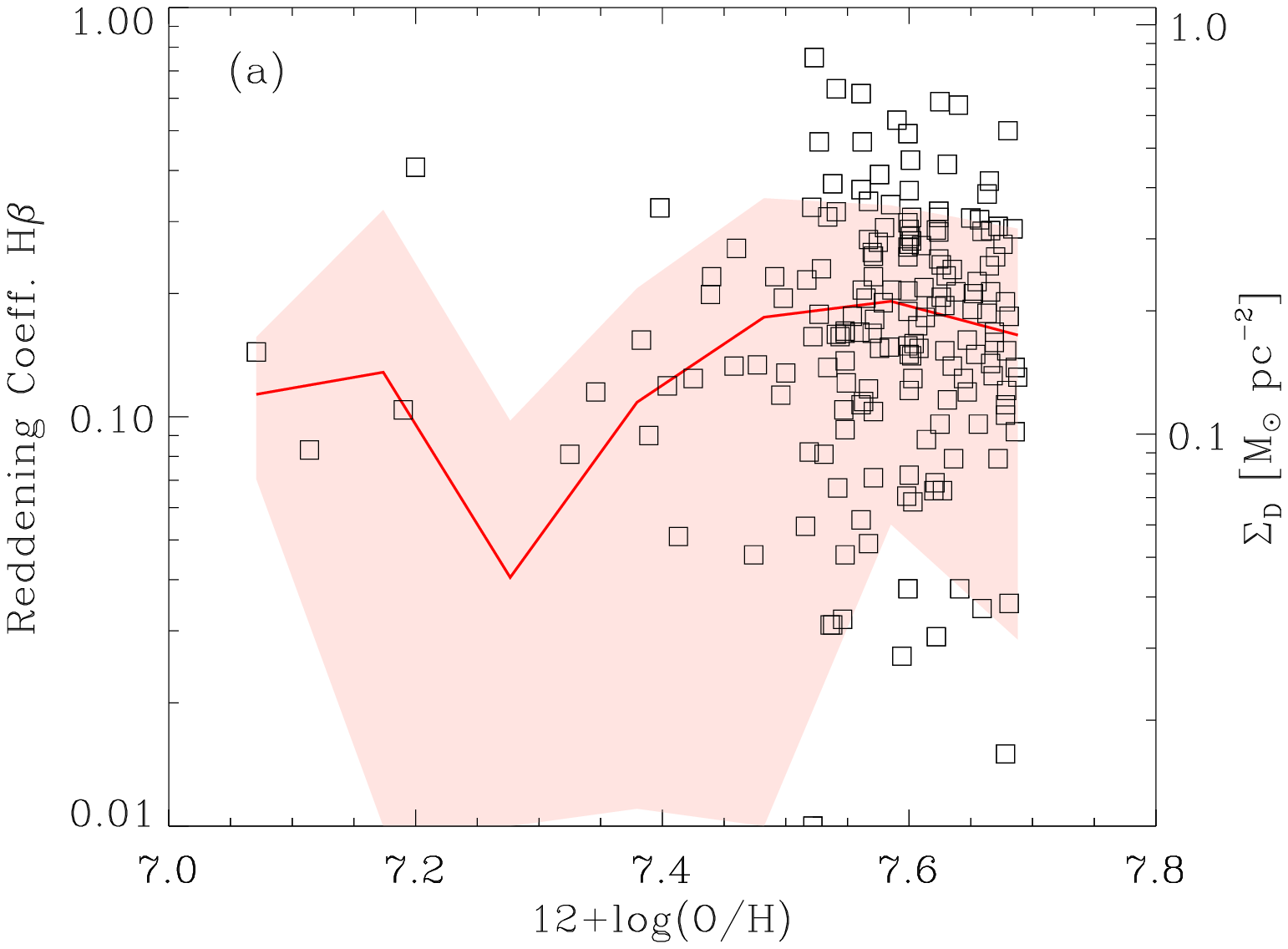}
\includegraphics[width=0.5\textwidth]{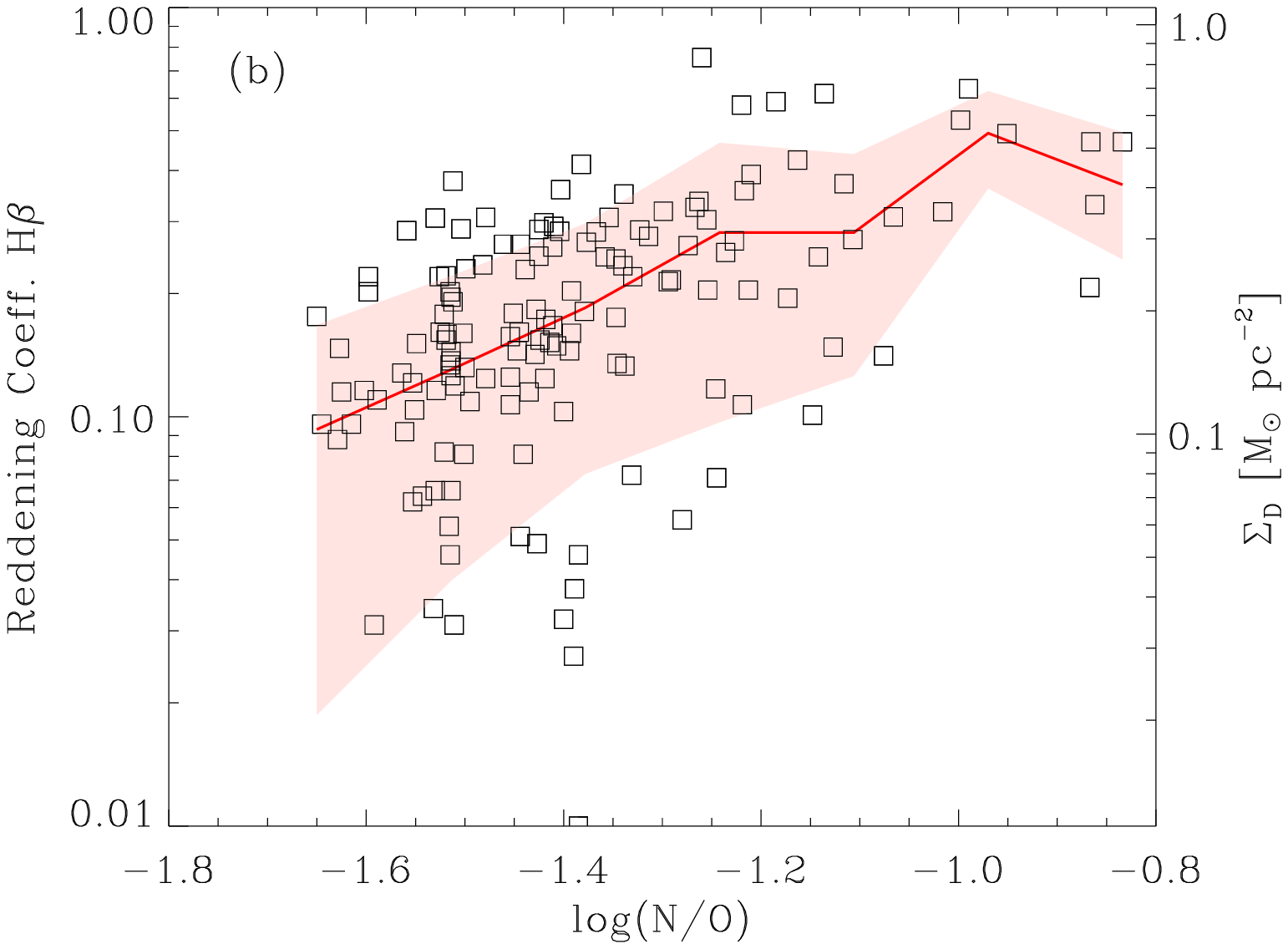}
\caption{
(a) Extinction coefficient at the wavelength of H$\beta$ versus 
oxygen abundance. The symbols show individual XMP
galaxies whereas the solid line and the shaded region correspond 
to the  mean and the standard deviation of the distribution of points.
There is no clear trend.
The second ordinate axis portrays the dust column density 
needed to produce the observed reddening (see text for 
details). 
(b) 
Extinction coefficient at H$\beta$  versus 
nitrogen over oxygen. The symbols show individual galaxies 
whereas the solid line and the shaded region correspond to the 
mean and the standard deviation of the distribution of points.
There is a trend, so that the reddening increases 
with increasing N/O.
}
\label{extinction}
\end{figure}

Figures~\ref{extinction}a and \ref{extinction}b show scatter plots of 
extinction  versus metallicity and N/O, respectively.
We find no clear relationship between extinction and metallicity 
(Fig.~\ref{extinction}a). However,  there is a hint of correlation
between extinction and log(N/O), so that the larger the
ratio between nitrogen and oxygen the larger the reddening
 (Fig.~\ref{extinction}b). The scatter of the possible relationship 
is large; the extinction coefficient increases 
only a factor of three over the full range of N/O,
and this increase is similar to the dispersion
of the individual galaxies in the scatter plot (Fig.~\ref{extinction}b).

The ordinate axes on the right of Figs.~\ref{extinction}a 
and \ref{extinction}b quantify the dust-mass column density responsible 
for the observed extinction. We have used Eq.~(\ref{dust_mass}) with a 
cross-section per unit mass $\langle\kappa_D\rangle=10^4\,{\rm cm^2\,g^{-1}}$,
which is representative of this quantity at H$\beta$  
\citep[e.g.,][Table~4]{2003ARA&A..41..241D}.
For the typical size of the giant HII regions 
in these objects, of the order of 150\,pc \citep[e.g.,][]{2013ApJ...767...74S,2015ApJ...810L..15S}, 
the mass in dust corresponding to the observed surface density
spans from 2$\times 10^2$ to 2$\times 10^4$\,${\rm M_\odot}$.
For reference, the dust mass of the prototypical XMP IZw\,18,
as inferred from modeling IR dust emission, is between 450 and 1800\,${\rm M_\odot}$
\citep{2014Natur.505..186F}. 
IZw18 is included in Table~\ref{main_list}, and 
its spectrum shows no extinction with an uncertainty of 0.07, which sets  an
upper limit of  1.5$\times 10^3$\,${\rm M_\odot}$ consistent with the value 
inferred from dust emission by \citeauthor{2014Natur.505..186F}

The distribution of extinction coefficients for the XMPs in Table~\ref{main_list}
has  a mean and a standard deviation given by 
\begin{equation}
\mu_{\rm c(H\beta)}\pm \sigma_{\rm c(H\beta)}=0.17\pm 0.14.
\label{extinction_mean}
\end{equation}
Using Eq.~(\ref{dust_mass}), this extinction corresponds to a dust mass
column density of, 
\begin{equation}
\mu_{\Sigma_D}\pm \sigma_{\Sigma_D}=(0.19\pm 0.15)\, {\rm M_\odot\,pc^{-2}}.
\end{equation}

%
\subsection{Stellar masses}\label{masses}

Stellar masses are usually estimated  from photometry via models that provide the
 light-to-mass ratio (M/L) given the observed colors of a galaxy 
\citep[e.g.,][]{2001ApJ...550..212B}.  This procedure is particularly uncertain for  
bursting XMP galaxies,
 with spectra having intense emission lines that are hard to capture by the model
used for M/L. However,  stellar mass is one of the central parameters characterizing 
galaxies, therefore, even with the above  caveat in mind, we have analyzed 
photometry-based stellar  masses of the XMP galaxies.
We use those from the  Max-Planck-Institute for Astrophysics -- Johns Hopkins University
(MPA-JHU)\footnote{{\tt http://www.mpa-garching.mpg.de/SDSS/DR7/}}
characterization of the SDSS-DR7 galaxy spectral data 
\citep[][]{2004MNRAS.351.1151B,2007ApJS..173..267S}.
Broad-band galaxy-integrated magnitudes from the SDSS-DR7 
database are  compared with a grid of theoretical 
galaxy spectra \citep{2003MNRAS.344.1000B} spanning a large range 
in star-formation histories \citep{,2007ApJS..173..267S}. 
From the difference between observed and theoretical magnitudes,
 a likelihood distribution for the mass of each 
galaxy is estimated. We use the median and the dispersion of this
 distribution in our analysis.

\begin{figure}
\includegraphics[width=0.5\textwidth]{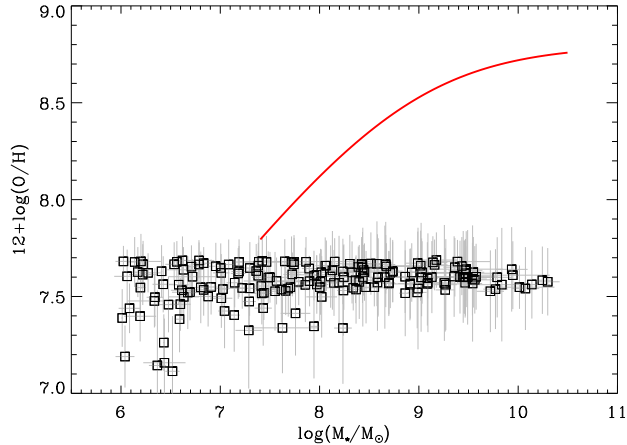}
\caption{
Metallicity vs stellar-mass scatter plot for the XMPs 
in Table~\ref{main_list} (the symbols with error bars). 
Although XMPs tend to be low-mass, there is a large spread in stellar 
masses.
When galaxies are chosen independently of metallicity, their
stellar masses and metallicities are related
--  the solid line represents the mean correlation 
observed and  parameterized by \citet[][]{2013ApJ...765..140A}.
There is no relationship in the case of the XMPs.}
\label{masmetal}f
\end{figure}

\begin{figure}
\includegraphics[width=0.5\textwidth]{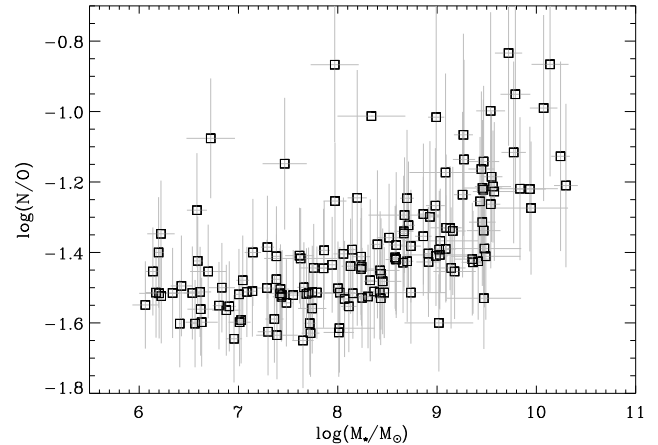}
\caption{
Scatter plot N/O vs stellar mass for the XMP galaxies in Table~\ref{main_list}
(symbols with error bars).  There is a clear tendency for N/O to increase
with increasing stellar mass when $\log({\rm M}_\star/{\rm M}_\odot) > 8.5$.
}
\label{masmetal2}
\end{figure}

\begin{figure}
\includegraphics[width=0.5\textwidth]{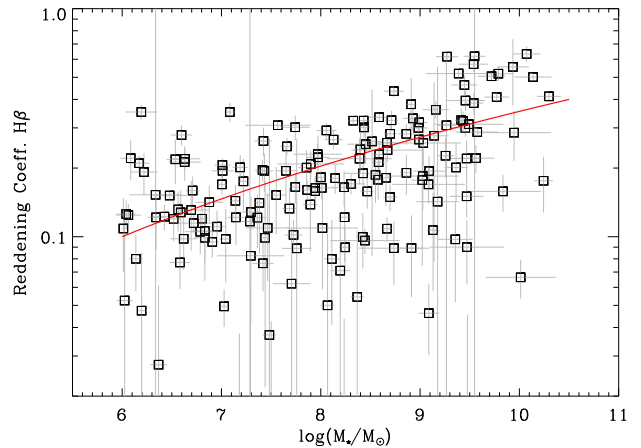}
\caption{
Scatter plot reddening coefficient  vs stellar mass for the XMP galaxies in Table~\ref{main_list}
(symbols with error bars).  There is a tendency for the reddening to increase
with increasing stellar mass, however, the spread of reddening for a given stellar mass is large.   
The straight line has been included to guide the eye. 
}
\label{masmetal3}
\end{figure}

The observed stellar masses have a mean and a standard deviation of
\begin{equation}
\mu_{\rm \log M_\star}\pm \sigma_{\rm \log M_\star}=8.0\pm 1.1, 
\label{mean_mass}
\end{equation}
where the masses have been expressed in solar masses. 
The XMPs are generally dwarfs, but there is a 
large spread in their possible masses -- 
${\rm \log(M_\star/M_\odot)}$ spans from 6 to 10.
This can be appreciated in Fig.~\ref{masmetal}, which shows
the scatter plot metallicity vs stellar mass.
We find no clear trend for the metallicity to vary with stellar mass.
This is forced by the selection criteria imposed on the XMPs.
In a randomly chosen set of galaxies, stellar masses and metallicities 
are correlated \citep[e.g.,][]{2004ApJ...613..898T}. 
Such relationship, as updated by  \citet[][]{2013ApJ...765..140A},
is included in Fig.~\ref{masmetal}. 
By construction, XMPs are restricted to have low metallicity and therefore 
cannot follow the general trend dictated by the red solid line in Fig.~\ref{masmetal}.  
However, hints of the underlying mass-metallicity relationship remains in the 
XMP sample, because
the lowest abundances are associated with galaxies in the low-mass part
of the mass distribution (Fig.~\ref{masmetal}). 
Figure~\ref{masmetal2} shows the scatter plot N/O vs stellar mass.
There is a clear trend for N/O to increase with increasing stellar mass 
for $\log({\rm M}_\star/{\rm M}_\odot) > 8.5$, which
is consistent with the relationship found in 
local galaxies \citep[e.g.,][]{2009MNRAS.398..949P,2013A&A...549A..25P}.   
In addition, some low-mass XMPs do show large N/O. The combination of low metallicity 
and high N/O can be understood if these galaxies suffered a major metal-poor gas 
accretion event that dropped O/H keeping N/O unchanged (see Sect.~\ref{discussion} for more details).
We also find a correlation between reddening and stellar mass (Fig.~\ref{masmetal3}).
It reflects the correlation between N/O and reddening (Fig.~\ref{extinction}b), 
given the fact that N/O increases with increasing stellar mass (Fig.~\ref{masmetal2}). 
We note that none of the large N/O targets have $12+\log({\rm O/H}) > 7.5$ (Fig.~\ref{distribution}a).

%
\subsection{Magnitudes and colors}\label{cmd}

\begin{figure}
\includegraphics[width=0.5\textwidth]{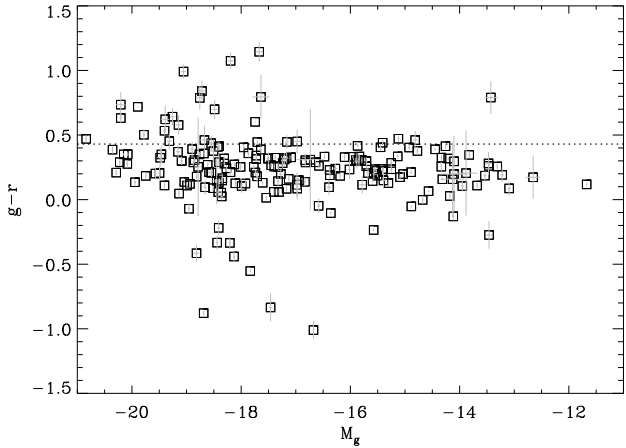}
\caption{
Color magnitude diagram of the XMPs.
The dotted line shows the color threshold to be a blue compact dwarf galaxy 
according to \citet{2003ApJS..147...29G}.
Most XMPs are bluer than this threshold, however, many of them 
are not dwarfs; often $M_g\leq -18$.  
}
\label{cmd_fig}
\end{figure}

Figure~\ref{cmd_fig} shows the color magnitude diagram in the color filters $g$ and $r$.
We use Petrosian magnitudes of the integrated galaxias as provided by SDSS.
The figure includes the color threshold to be a blue compact dwarf galaxy as
defined by \citet{2003ApJS..147...29G}
\citep[and transformed from U and B to $g$ and $r$ by][]{2008ApJ...685..194S}.
Most XMPs lie below the threshold and thus look blue in broad-band colors.
However, a significant part of them are not dwarf, with absolute magnitude 
in excess of $M_g\simeq -18$. The fact that some of them are  not dwarfs is also inferred 
from the distribution of masses in Fig.~\ref{masmetal}, which contains 
galaxies having masses  larger than $10^9\, {\rm M}_\odot$.  These high-mass XMPs often correspond to the most
distant objects, with redshifts above 0.05 and distances larger than 220 Mpc. 

%
\subsection{Number density of XMPs in the local Universe}\label{number_density}

The SDSS spectroscopic legacy sample  was designed
as a  magnitude-limited survey. Therefore, 
it is relatively simple to correct for the luminosity~bias 
(i.e., the Malmquist~bias) and so, to compute  
the volume density of particular objects. In \paperi\  we carried out
an estimate of the number density of XMPs in the local Universe.
Since the number of XMPs has increased substantially from \paperi, 
such an estimate has become obsolete, and is updated here. 

We employ the so-called $V_{max}$ approximation by 
\citet{1968ApJ...151..393S} used to, e.g., determine luminosity 
functions of galaxies \citep[][]{2000ApJS..129....1T}. The number density
of galaxies with the property $X$, $n(X)$, is just
\begin{equation}
n(X)=\sum_i{1\over {V_i}},
\label{pi_eq}
\end{equation}
where the sum includes all galaxies in the sample
with property $X$, and $V_i$ represents the maximum volume in 
which the $i$-th galaxy of the sample could be observed. 
In a magnitude-limited sample, so that all galaxies brighter than the 
apparent magnitude $m_{lim}$ are included,
\begin{equation}
V_i={{d_i^3}\over{3}}\Omega,
\label{pi_eq7}
\end{equation}
with $\Omega$ the solid angle covered by the survey, and 
$d_i$ the maximum distance at which the $i-$th galaxy can
be observed,
\begin{equation}
\log(d_{i}/l_i)= {{1}\over{5}}(m_{lim}-m_i).
\label{pi_eq8}
\end{equation}
The maximum distance depends on the absolute magnitude of the galaxy,
which can we written in terms of the distance to the 
galaxy, $l_i$, and the apparent magnitude, $m_i$, as it has been done for  
Eq.~(\ref{pi_eq8}). 

A few caveats are in order before applying 
Eq.~(\ref{pi_eq}) to the dataset. 
SDSS is not truly magnitude limited for a number of reasons.
Some bright galaxies are not observed in crowded field due to
fiber collision \citep[see][]{2002AJ....123..485S}. Low surface brightness galaxies 
tend to be missed even if they have significant integrated magnitudes
 \citep[e.g.,][]{2005ApJ...631..208B,2015MNRAS.448.2687J}. 
The actual galaxy catalog sample also contain objects coming from
complementary  SDSS searches, which often are fainter than the 
magnitude limit of the main survey. 
The first problem is of no relevance since most observed 
XMP galaxies are isolated \citep[][]{2015ApJ...802...82F}.
The second problem is likely more serious since very faint galaxies
tend to have low surface brightness as well \citep{1999ASPC..170..169S}.
Quiescent
XMP galaxies like Leo P \citep{2013AJ....146....3S} or KJ~78 
\citep{2015MNRAS.448.2687J} show a surface brightness around 
24~mag\,arsec$^{-2}$, and objects like them would hardly 
appear in our selection since SDSS is almost blind at this low surface 
brightness \citep{2005ApJ...631..208B}. As it is discussed in the introduction 
of the paper, 
we bypass the problem dividing  XMPs into quiescent and 
bursting. Our search in SDSS is sensitive to the bursting XMPs 
only, and so are the properties derived here. This also holds for the  number density to 
be inferred assuming a magnitude limited sample. 
The third caveat, namely the existence of XMPs from complementary 
searches, is also  
an issue since a significant part of our XMPs 
belong to these additional catalogs. We solve this problem by applying 
Eq.~(\ref{pi_eq8}) only to those XMPs with $m_i < m_{lim}$. Using
$m_{lim}=17.8$ in the filter $r$, then Eqs.~(\ref{pi_eq}), (\ref{pi_eq7}), 
and (\ref{pi_eq8}) render
\begin{equation}
n\big({\rm O/H} \leq {1\over {10}}\,{\rm O/H}_\odot)= 
(3.4\pm 0.9)\,\times 10^{-3}\,{\rm Mpc^{-3}},
\label{actual_density}
\end{equation}
with the error bar considering only the poissonian error associated
with the process of galaxy counting.

The density in equation~(\ref{actual_density}) is larger than the density obtained 
from the sample in \paperi, which is consistent with 
the fact that the present work contains almost ten times more targets.
In terms of the total number of galaxies, XMPs represent some 2\,\% of
all galaxies in a given local volume. The number density of galaxies 
in the local universe, $\sim 0.17\,{\rm Mpc^{-3}}$, has been taken from the 
normalization of the total $r$-band luminosity function of SDSS galaxies
 by \citet{2005ApJ...631..208B},
which includes galaxies brighter than absolute magnitude -12.
Even with a significant increase
with respect to the previous estimate in \paperi, XMP galaxies are still 
rare.

%
%
\subsection{Low-metallicity threshold}\label{threshold}

One of the puzzling properties of XMP galaxies is the existence 
of a threshold for the lowest metallicity. This threshold is also present
in the XMPs revealed here. Figures~\ref{distribution}a and \ref{distribution}b 
show that 
\begin{equation}
12+\log({\rm O/H}) \gtrsim 7.1,
\end{equation}
which approximately corresponds to 1/40 times the oxygen abundance in the 
solar composition. This value is similar to, although slightly higher than,  
the record-breaking  low-abundance galaxy SBS0335-052W 
\citep[$12+\log({\rm O/H})\simeq 7.0$, corresponding to 1/50th of the solar 
composition;][]{2009A&A...503...61I}.

The reason why such a threshold exists is still a riddle.
The metallicity of the gas produced during the Big Bang is virtually 
metal-free, and population III stars contaminate the inter-galactic 
medium with a metallicity lower than $10^{-4}$ times the solar metallicity
\citep[e.g.,][]{2004ARA&A..42...79B}.
One can envisage that some of the galaxies formed from this 
pristine gas have been so dull forming stars that they still 
retain part of the original gas. Why none of these {
pristine galaxies is producing stars at present? 
Several explanations have been offered in the literature, going from  
the self-enrichment of the HII region used for measuring, to  
technical difficulties for metallicity determinations below the threshold
\citep[see][]{2011EAS....48...95K}.
One additional possibility is offered by the cosmic-web gas-accretion scenario,
where the starbursts in XMPs are triggered by gas recently accreted 
from the cosmic web \citep[see ][and references therein]{2014A&ARv..22...71S}. 
Numerical simulations predict the cosmic web gas to accumulate metals from the 
metal-enriched outflows of dwarf galaxies. These metals add up along the Hubble time 
so that at redshift zero the cosmic-web metallicity is predicted to have a value close
to the observed threshold \citep[e.g.,][]{2012MNRAS.420..829O}. 
Should this explanation were correct, then the observational lower limit is tracing the 
metal content of the cosmic web at present.

%
\subsection{Morphology}\label{morphology}

\citet{2008A&A...491..113P}, \paperi , and later \citet{2013A&A...558A..18F}  pointed
out that bursting XMPs 
tend to be cometary or tadpole-like. Such association between 
metallicity and morphology is notable since metallicity is a property  dictated 
by the spectrum and, therefore, it implies that the spectrum conditions 
the morphology of the galaxy. 
In order to check whether the new sample of XMPs maintains this 
preference for the cometary shape, we repeated the eyeball classification
carried out by  \citet{2013A&A...558A..18F}, which was based on the
scheme presented in \paperi . The galaxies may be (1) {\em symmetric}
if they present a centrally concentrated emission with no obvious 
structure in the outskirts,  (2) {\em cometary}  for a head-tail structure with 
an identifiable knot at the head, (3) {\em 2-knot}  for a structure with two knots,
and (4) {\em multi-knot} for a diffuse structure with multiple star-formation knots,
including irregular galaxies. 
Figure~\ref{classes_mph} illustrates these classes with two examples of 
each galaxy morphology.
\begin{figure*}
\includegraphics[clip,width=4.3cm]{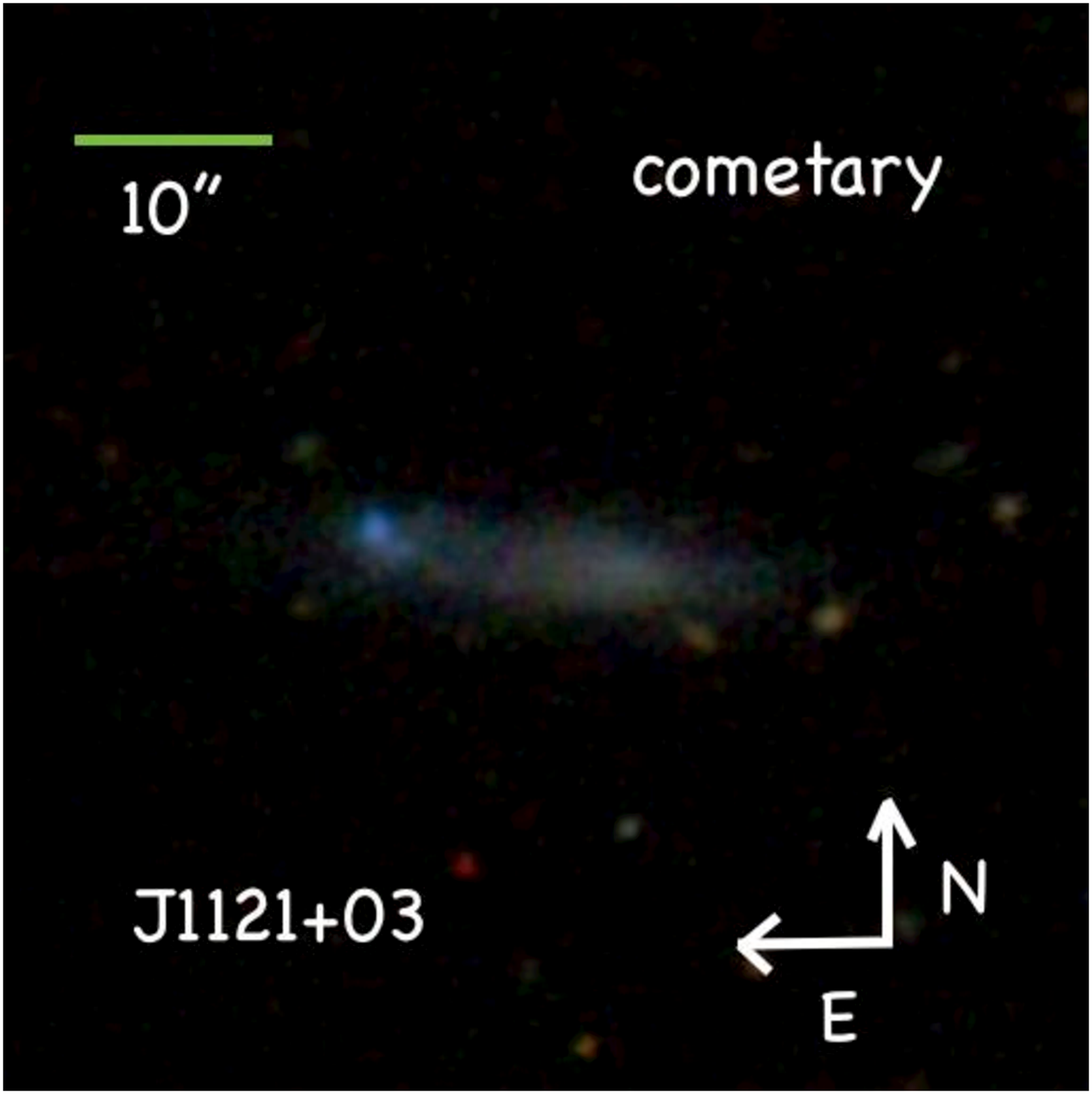}\hspace{0.12cm}
\includegraphics[clip,width=4.3cm]{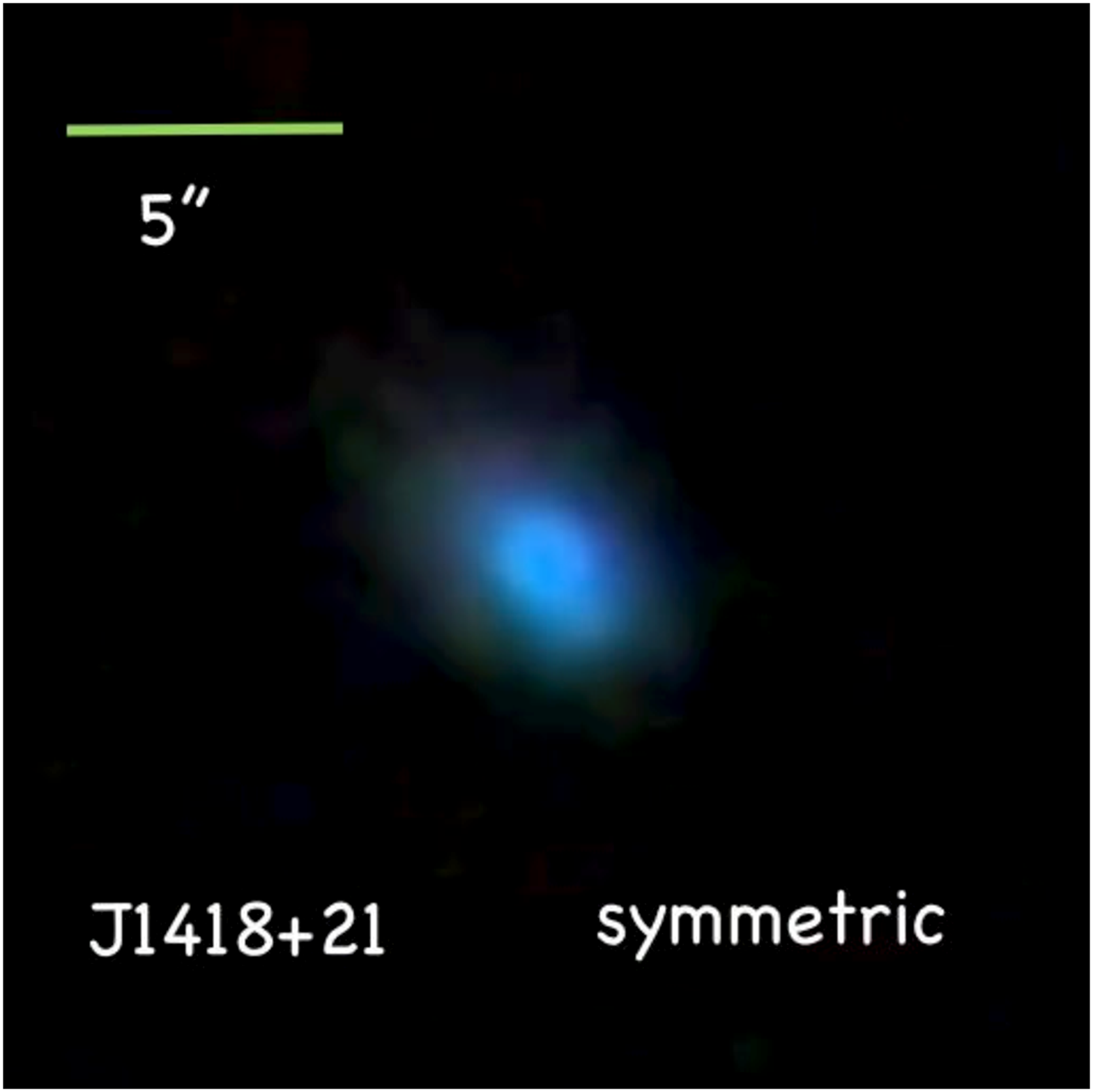}\hspace{0.10cm}
\includegraphics[clip,width=4.3cm]{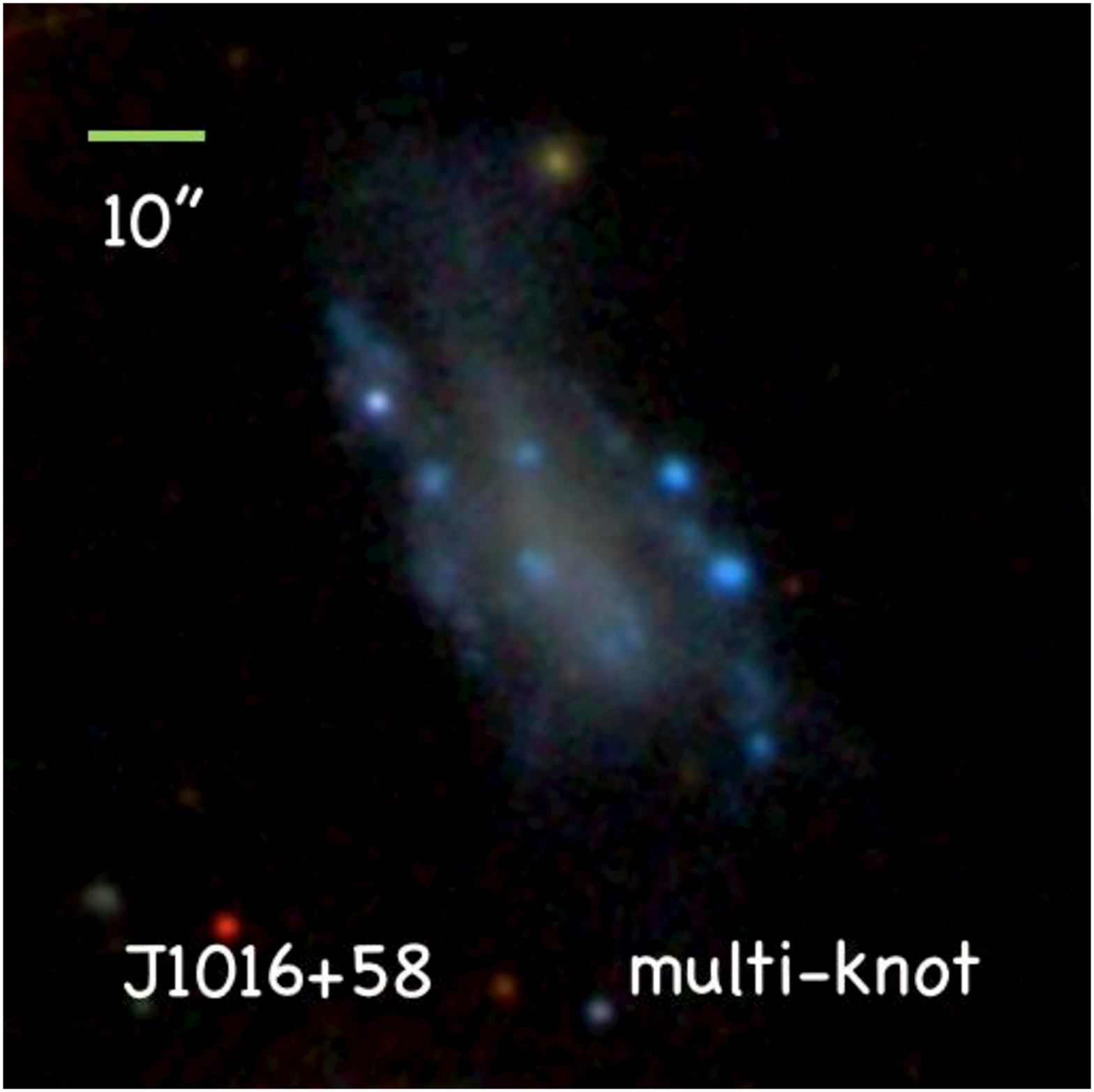}\hspace{0.10cm}
\includegraphics[clip,width=4.3cm]{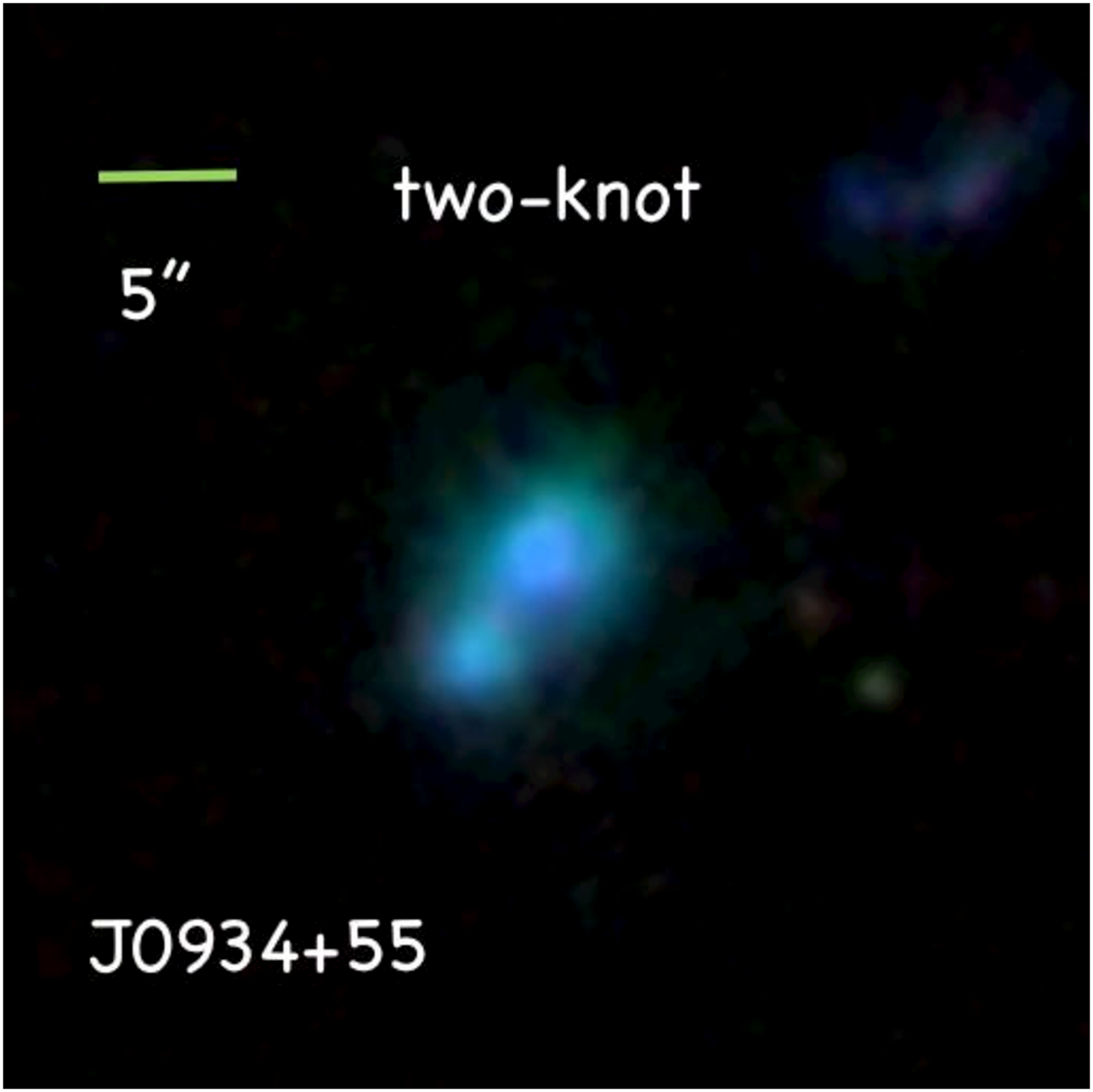}\\
\vspace{-2.mm}\\
\includegraphics[clip,width=4.3cm]{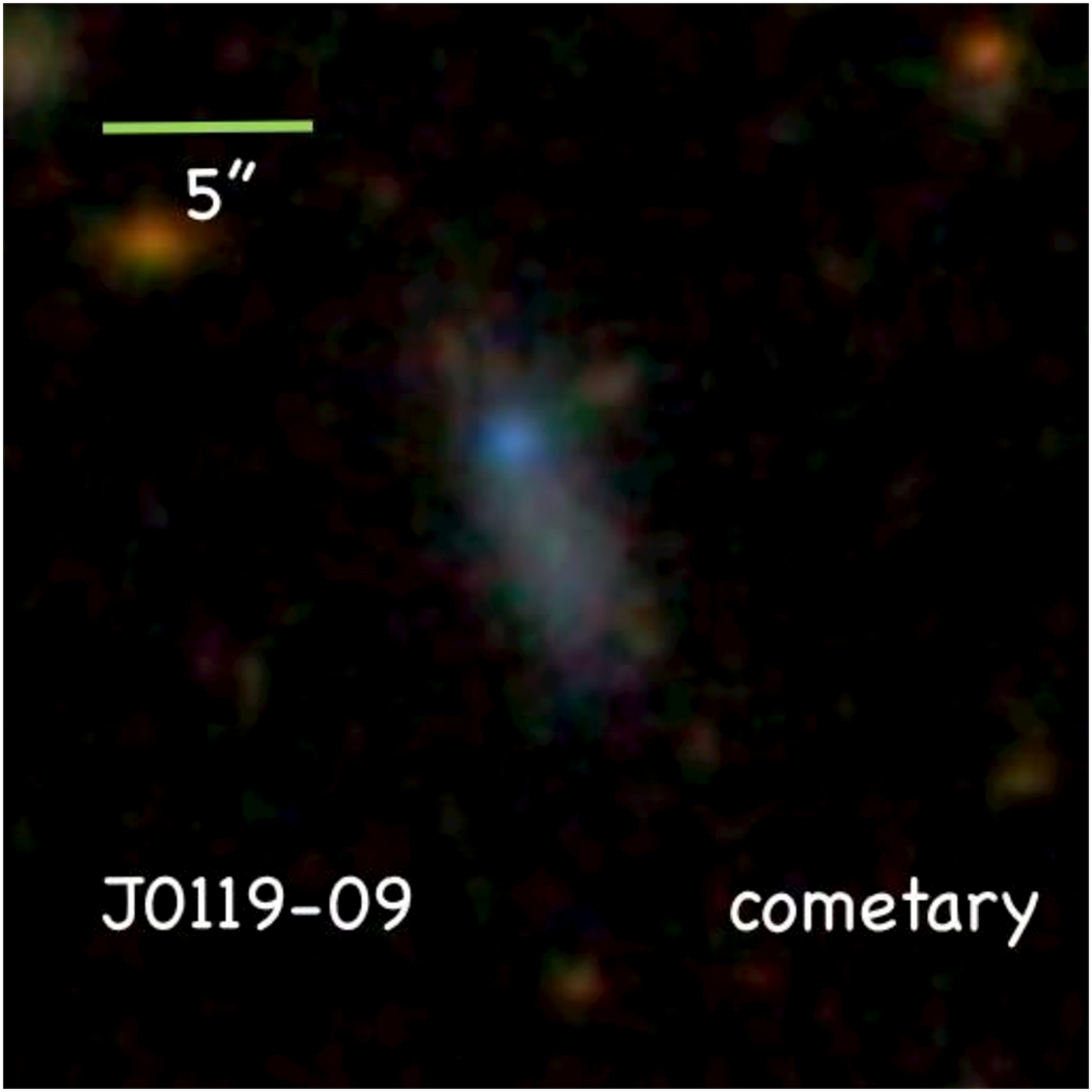}\hspace{0.05cm}
\includegraphics[clip,width=4.3cm]{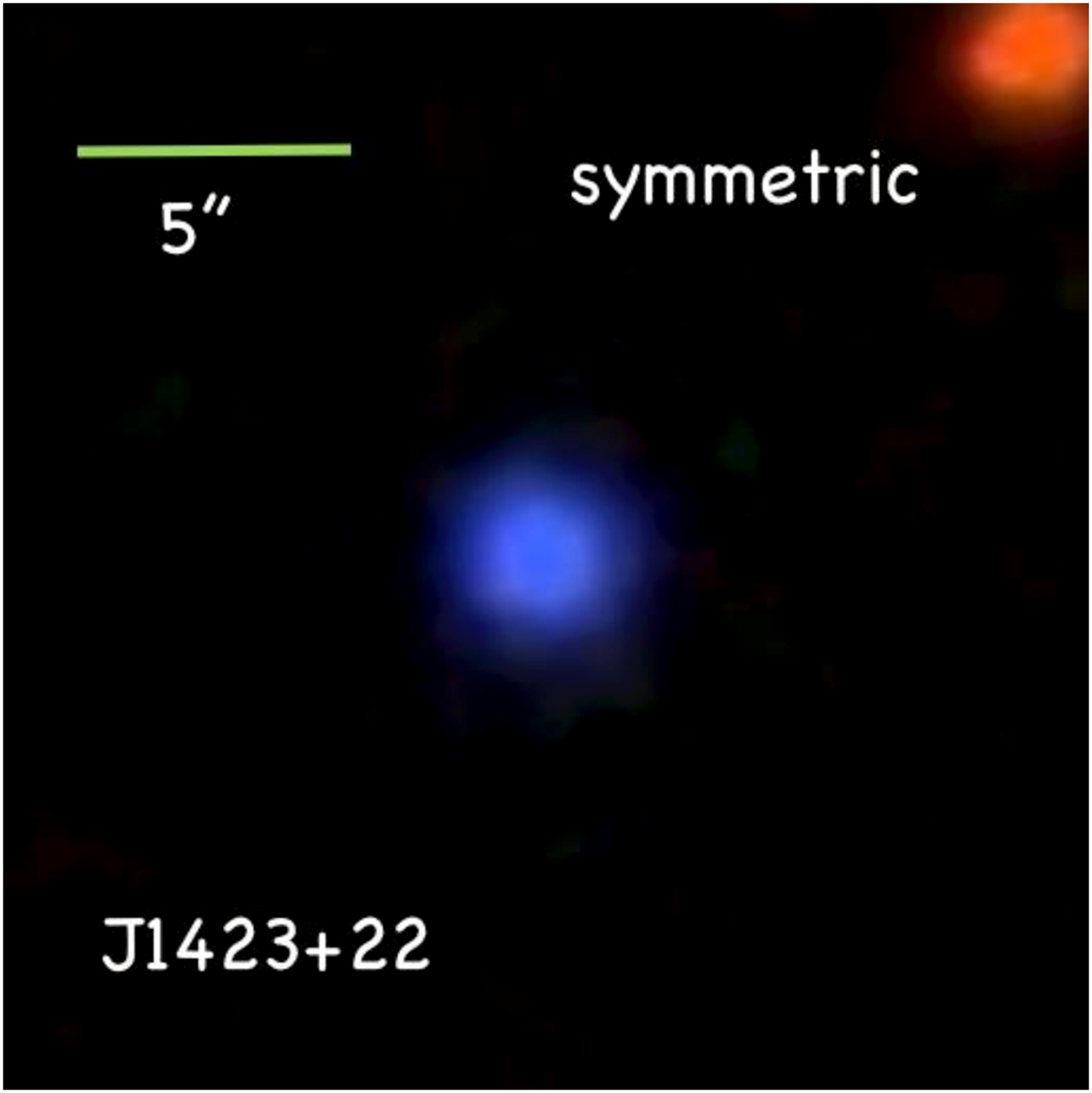}\hspace{0.08cm}
\includegraphics[clip,width=4.3cm]{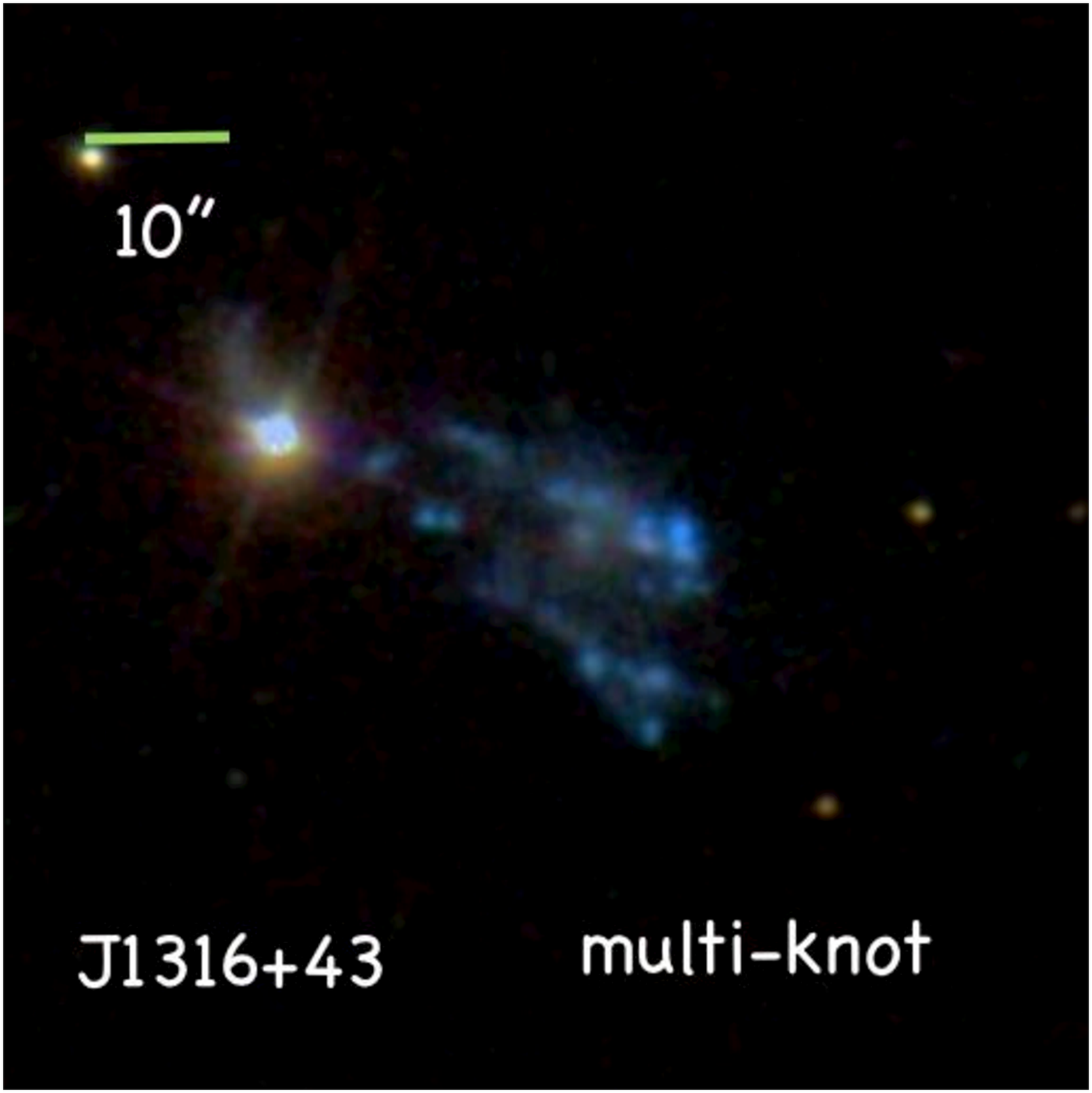}\hspace{0.10cm}
\includegraphics[clip,width=4.3cm]{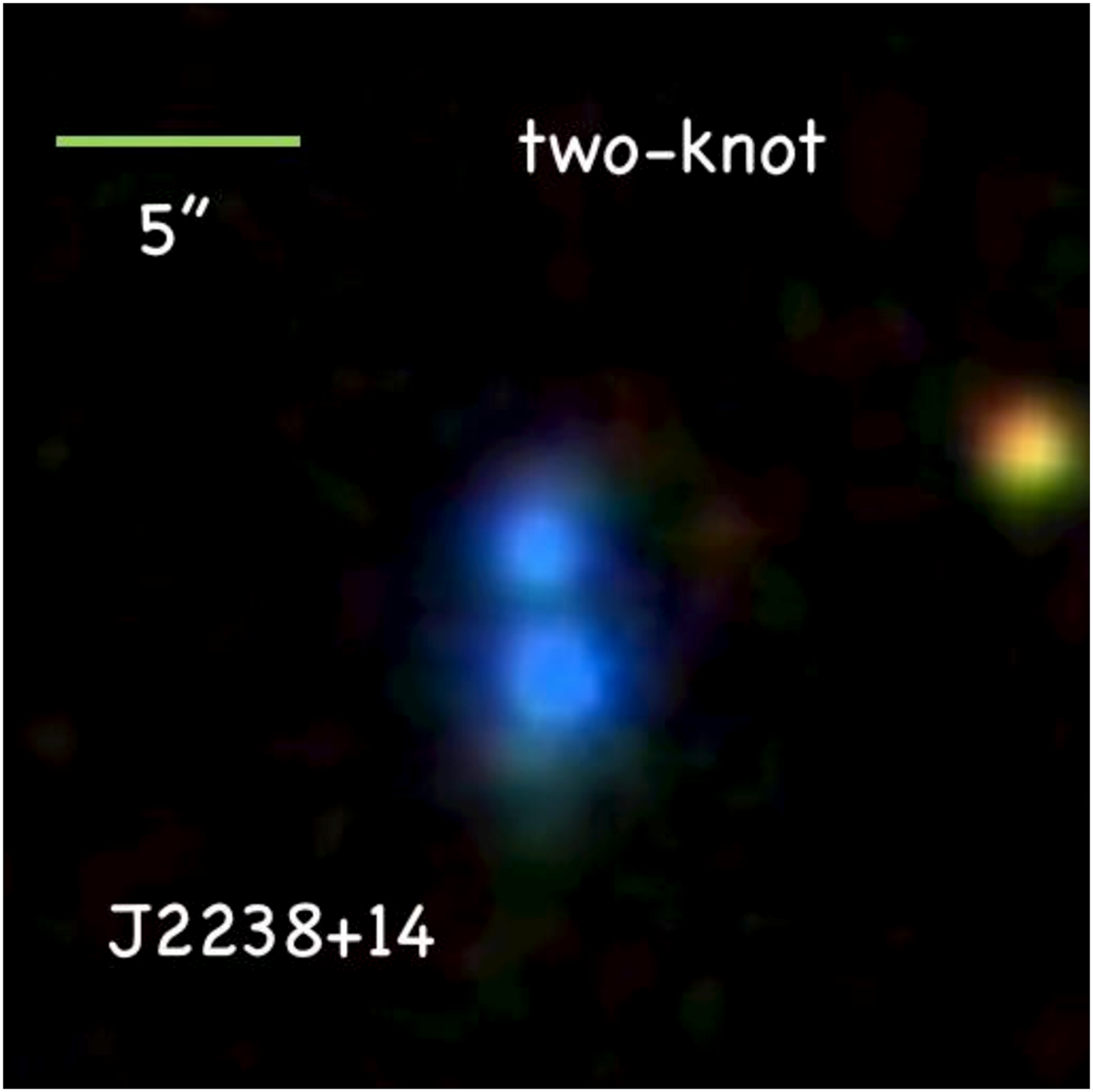}\\
\caption{Examples of the various morphological types found 
among the XMPs; from left to right cometary, symmetric, multi-knot, and 
two-knot (see main text for details). The bar on top of each image 
gives an angular scale on the sky. The image on the top-left panel also 
points out north and east, which are common to all 
images. Galaxy images are from SDSS-DR12. The two-knot example on 
the upper row corresponds to IZw18.}
\label{classes_mph}
\end{figure*}
We use color images from SDSS for the visual inspection, and 
the result of the classification turns out to be,
\begin{equation}
\begin{tabular}{lr}
Cometary&57\,\%,\\
Symmetric&23\,\%,\\
Multi-knot&10\,\%,\\
2-knot&4\,\%,\\
others&6\,\%.
\end{tabular}
\label{morpho}
\end{equation}
In agreement with previous estimates, cometary happens to be the dominant shape.
Asymmetric shapes (i.e., cometary plus 2-knot plus multi-knot) represent 71\,\% of the
sample. These results are in quantitative agreement with \paperi , where 75\% of the targets 
are asymmetric, and with \citet{2013A&A...558A..18F}, where 
cometary, multi-knot and 2-knot represented 52\,\% , 16\%, and 10\% of the XMP sample, 
respectively.
Perhaps the main difference is a slight excess of symmetric targets in the present
sample, which we attribute to insufficient spatial resolution since our targets 
have larger redshifts than the other samples (Sect.~\ref{cmd}), thus complicating
the detection of sub-structure. In addition, this time we include a 5th class
to the classification ({\em others} in Eq.~[\ref{morpho}]) to collect objects that 
do not fit in the main categories easily (e.g., mergers, or elongated galaxies
without an identifiable knot).
In addition, two of the objects are HII regions in the outskirts 
of large spirals (\#\,20 and 59).  

We note that a significant part of the symmetric objects looks
like the {\em green peas} first described by 
\citet{2009MNRAS.399.1191C}. One example is shown in Fig.~\ref{classes_mph} --
the blue symmetric galaxy J1423+22.
Depending on the redshift, green-pea like galaxies may look green, pink or blue in SDSS
images \citep{2013RMxAC..42..111S}. The identification of green peas with
metal-poor galaxies was carried out by  
\citet{2010ApJ...715L.128A,2012ApJ...749..185A},
and their rounded morphology seems to be associated 
with limitations in spatial resolution rather than being
intrinsic \citep[e.g.,][]{2014A&A...568L...8A}.
The finding of green peas in our sample is encouraging since 
it naturally connects local XMPs to those expected at higher 
redshifts, which will appear in surveys deeper than SDSS-DR7 
(e.g., eBOSS\footnote{eBOSS stands for extended baryon oscillation spectroscopic survey,
and a description can be found in {\tt http://www.sdss.org/surveys/eboss/}.
} within SDSS IV).

%
\subsection{Underlying stellar populations}\label{agess}

\begin{figure}
\includegraphics[width=0.5\textwidth]{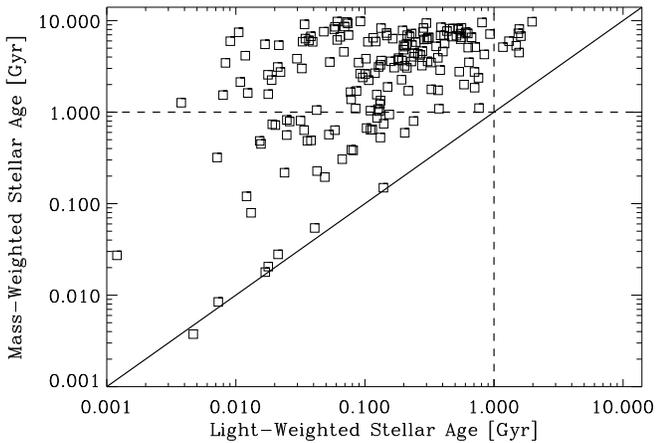}
\caption{
Mass-weighted stellar age vs light-weighted stellar age for the 
XMPs, both derived using {\sc starlight} (Sect.~\ref{fluxes}). 
The light-weighted mean is biased toward the younger 
stellar populations, whereas the mass-weighted mean is not.
The vertical dashed line shows a  light-weighted stellar age of 1 Gyr.
Note that most XMPs are to the left of this threshold and thus
present a stellar spectrum typical of young stellar populations.
The horizontal dashed line shows a  mass-weighted stellar age of 
1 Gyr. Most XMPs are above this line implying that, despite their 
seemingly young spectrum, XMPs also contain aged stellar 
populations -- often 10\,Gyr old.
The ages in the axes are given in Gyr, and 
the slanted solid line corresponds to the one-to-one 
relationship.
}
\label{ages}
\end{figure}
In order to decontaminate the emission-line fluxes from stellar 
absorption,  we fitted a composite stellar spectrum to the observed 
continuum (Sect.~\ref{fluxes}). As byproduct of this exercise, 
we have information on the stellar populations co-existing with the 
gas responsible for the emission line spectrum. The stellar-population 
decomposition is not free from uncertainties and degeneracies,
however, it suffices to provide a general view of the stellar
properties \citep[e.g.,][]{2005MNRAS.358..363C,2014A&A...561A.130C}.  

We find that the underlying stellar populations are generally young, 
although in almost all cases there are also old stellar components
contributing to the observed spectrum. This can be appreciated in 
Fig.~\ref{ages}, which shows the mass-weighted stellar age
versus the light-weighted stellar age for the set of XMPs. 
The light-weighted mean is strongly biased toward young 
stars, and we find that most light-weighted ages are younger than 
1\,Gyr (i.e., they are to the left of the vertical dashed line in 
Fig.~\ref{ages}). The 
mass-weighted mean,
however, does not suffer from this bias, and most XMPs have 
mass-weighted ages significantly larger than 1\,Gyr (i.e., they
are above the horizontal  dashed line in Fig.~\ref{ages}). In other words,
most XMPs seem to have been forming stars not only at present
but also during the last Gyr. On top of this recent star-formation, most
of them contain old stars, often as old as 10\,Gyr or so.  
The fact that the class of XMP galaxies contains evolved stellar 
populations is by no means unknown (see Sect.~\ref{introduction}). 
Their large difference between mass-weighted age and 
light-weighted age is common to most local star-forming 
galaxies  \citep[e.g.,][]{2010MNRAS.404.2037P,2012ApJ...756..163S}.


\subsection{Large scale environment}\label{environment}

As it was done by \citet{2015ApJ...802...82F} for the XMPs in \paperi ,
we have explored the large scale environment of the new list of 
bursting XMPs
using constrained N-body cosmological numerical simulations of
the local universe \citep{2014MNRAS.445..988N}.  
The cosmological parameters provide a statistical description of the 
Universe. The constrained simulation that we use 
selects a particular realization of the current cosmological model 
that reproduces the spatial distribution of galaxies in the local universe as observed  by 
2MRS\footnote{The 2MASS Redshift Survey (2MRS) is the spectroscopic follow-up 
of the two micron all sky survey 2MASS which provides a 3D distribution 
of local galaxies -- RA, DEC and redshift.} 
\citep[][]{2012ApJS..199...26H}. 
The method is based on a self-consistent Bayesian machine-learning
algorithm, and it provides a description of the local cosmic web
dark matter with a resolution around 2~Mpc in a box of 180~Mpc$\,h^{-1}$ side
\citep{2012MNRAS.427L..35K, 2013MNRAS.435.2065H, 2013MNRAS.429L..84K}.
This simulation, tuned to replicate the local distribution of galaxies,
tell us whether the gravitational field at each point forces the matter to collapse 
in three directions (knot), in two directions (filament), in one direction (sheet),
or if the matter is expanding in all three directions (void). This classification,
based on the local gravitational potential, correlates very 
well with the local dark-matter overdensity, which increases from voids to knots.
The tendency of a particular type of galaxy to prefer one of the four 
environments is quantified in terms of the excess probability ratio $\eta(\tau,\epsilon)$,
defined as the ratio between the probability that a galaxy of type $\tau$ 
appears in environment $\epsilon$, and the probability of having 
any type of galaxy in this environment. The excess probability 
for the XMPs is shown in Fig.~\ref{constrained}. For reference,
the figure also includes the excess probability for ellipticals (E),
lenticulars (SO), spirals (Sp) and irregulars (Irr) as worked out 
from 2MRS galaxies by \citet{2014MNRAS.445..988N}.
Clearly, XMPs have a strong tendency to appear in voids and 
to avoid knots. This tendency is more clear in XMPs than in 
any of the other galaxy types.
\begin{figure*}
\includegraphics[width=0.5\textwidth]{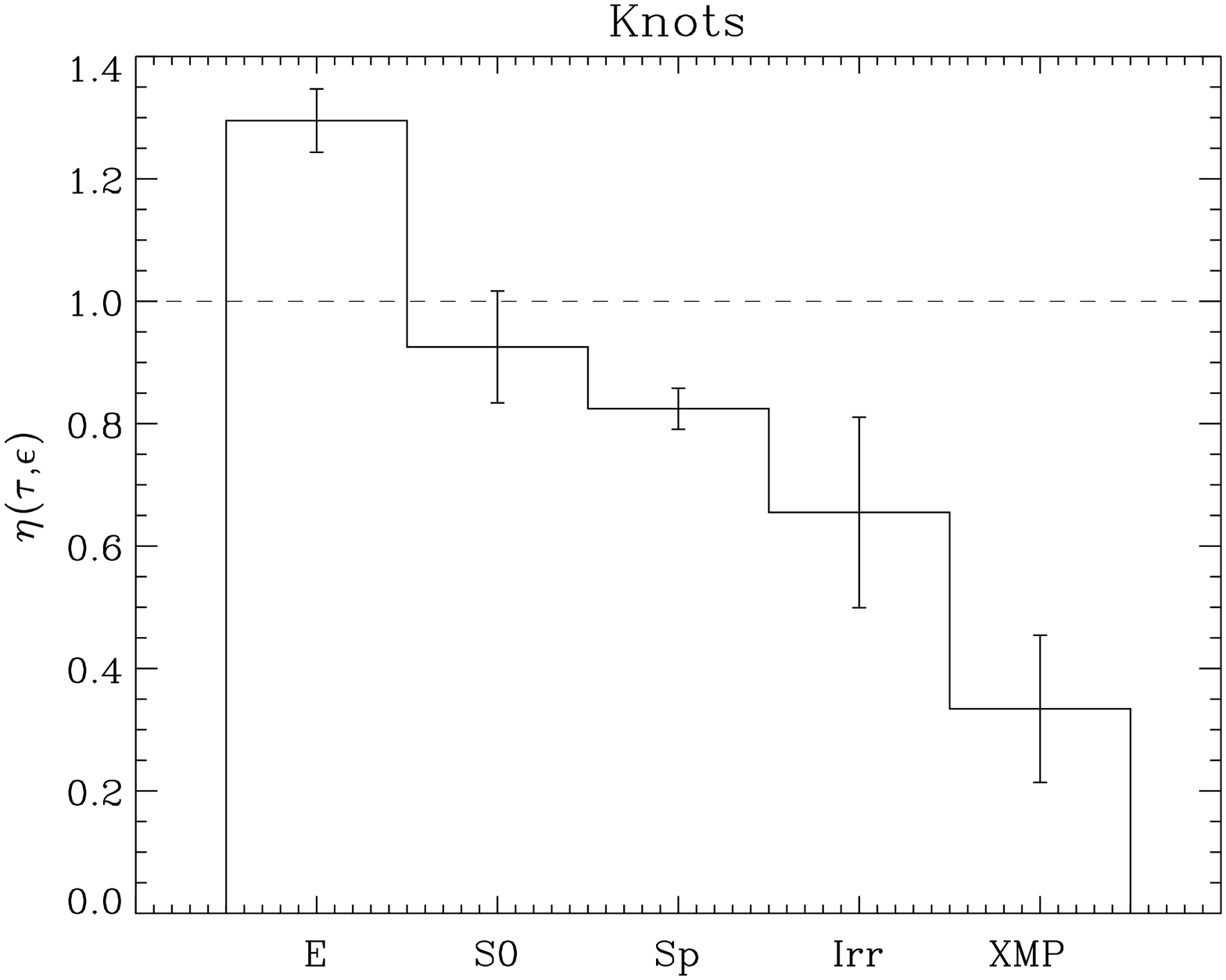}
\includegraphics[width=0.5\textwidth]{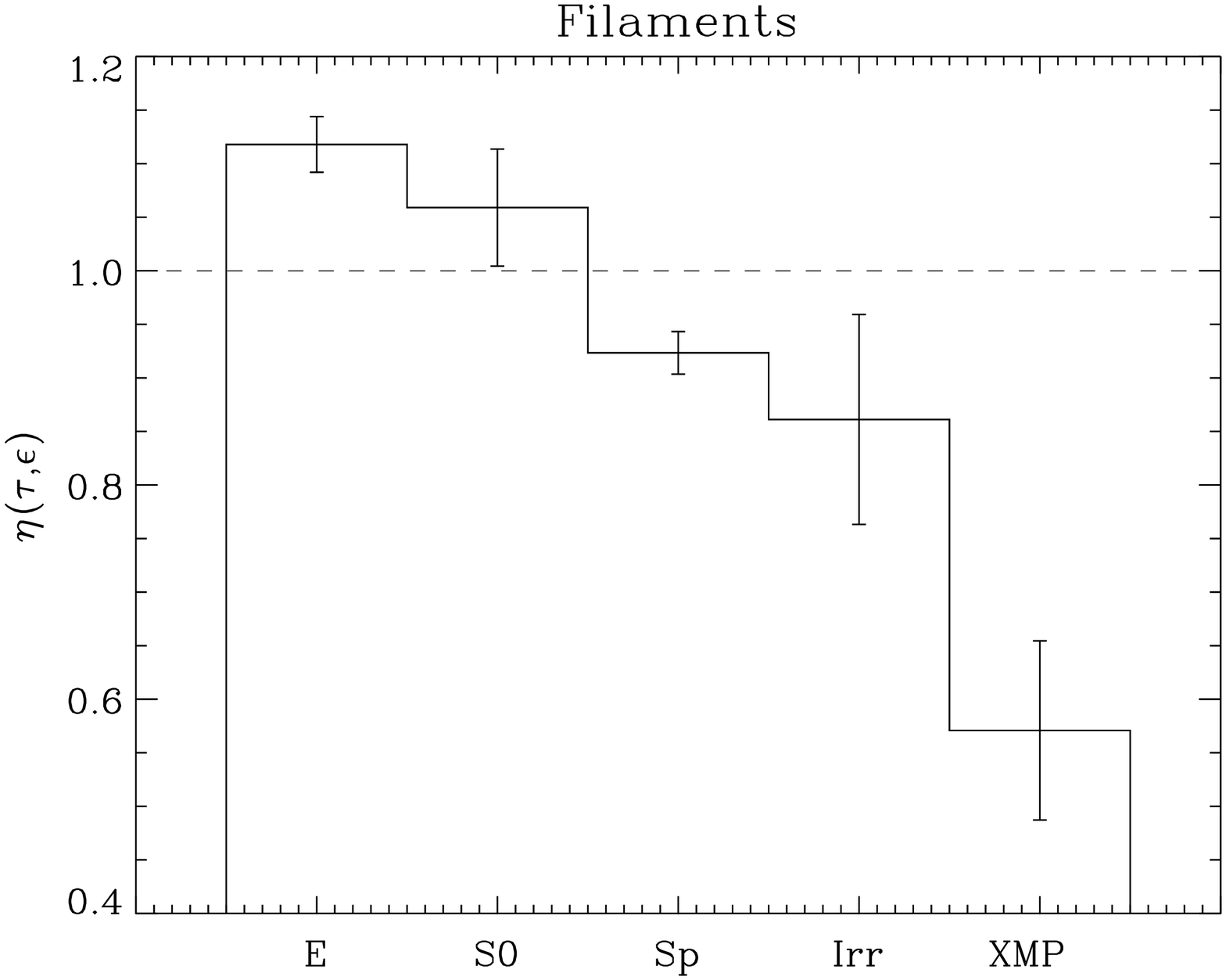}\\
\includegraphics[width=0.5\textwidth]{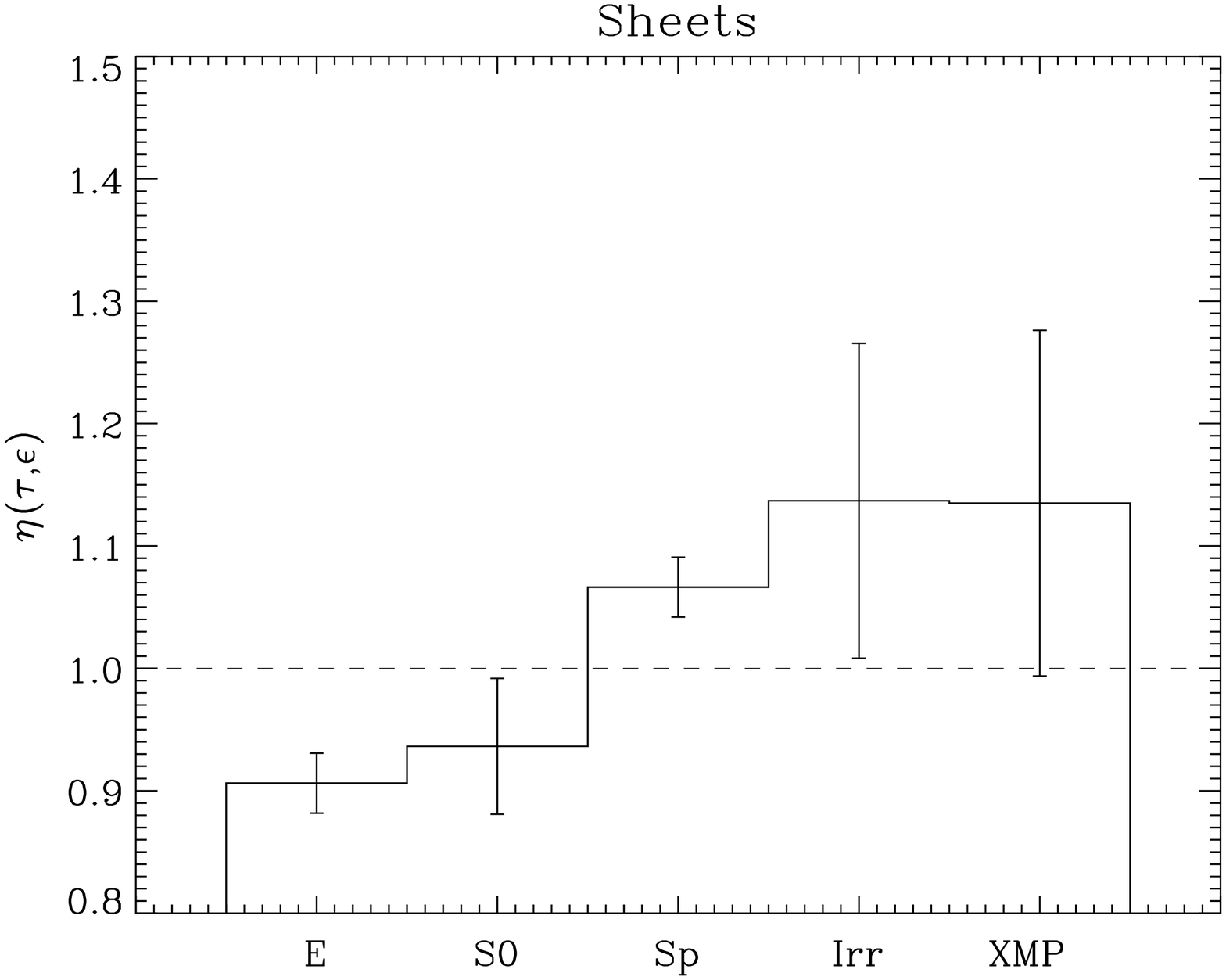}
\includegraphics[width=0.5\textwidth]{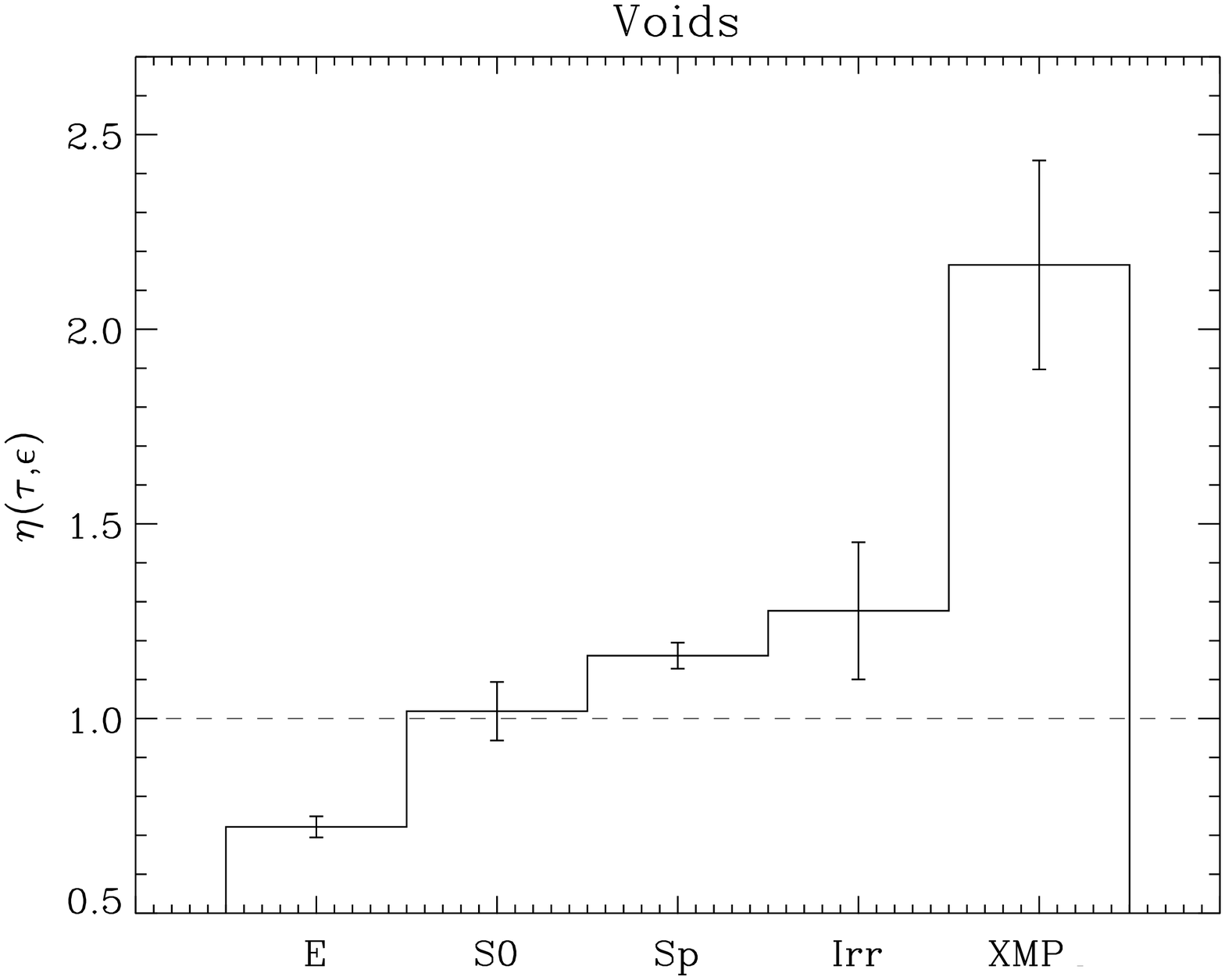}
\caption{Excess probability ratio for the XMPs.  E, S0, Sp, and Irr galaxies
are also included for reference (see main text for details). 
The four panels correspond to the four types of environments: 
voids, sheets,  filaments, and knots. XMPs have a strong tendency to appear in voids and 
to avoid knots.}
\label{constrained}
\end{figure*}
These results are very similar to those found by \citet{2015ApJ...802...82F}
for the XMPs in  \paperi .

%
\section{Discussion and conclusions}\label{discussion}

Due to their astrophysical interest and rareness (Sect.~\ref{introduction}), we have 
carried  out a  systematic search for XMP galaxies in SDSS-DR7. 
Starting from all 
galaxies with spectra having redshift smaller than 0.25,
we end up with a list of only \nxmps\ XMPs (Table~\ref{main_list}). 
We select galaxies having [OIII]$\lambda$4363 particularly intense
with respect to [OIII]$\lambda$4959 and [OIII]$\lambda$5007.
As it is explained in Sect.~\ref{kmeanssearch}, this implies having 
a high temperature gas and, since metals are efficient coolants, it is
the fingerprint of low-metallicity gas.  A first automated search was carried 
out using k-means to classify all galaxy spectra according to 
the shape  in the region that contains  [OIII]$\lambda$4363
and [OIII]$\lambda\lambda$4959,5007. The  \ncandidates\ objects 
in the classes of high [OIII]$\lambda$4363 underwent
a closer scrutiny by
computing their O abundance using the method by 
\citet[][Sect.~\ref{hcm}]{2014MNRAS.441.2663P}, which is a model-based 
approach consistent with the direct method. Those with abundance 
smaller than one tenth of the solar value ended up in the 
final list (Table~\ref{main_list}). The existence of 
uncertainty in the metallicity determination, typically 0.17\,dex, implies that some of the
XMPs are false positives, and some of the discarded  targets are
false negatives (Appendix~\ref{appendix}). 
We offer a second list with \nxmpsdos\ XMP 
candidates (Table~\ref{candidates}), selected under less restrictive 
noise conditions, which contains objects that upon refinement of 
the metallicity estimate may turn out to be XMPs. The search in the
present paper is complementary to the one carried out in \paperi\ , 
where k-means is used in the spectral region around H$\alpha$, 
where the metal-poor signpost is the weakening of 
[NII]$\lambda$6549 and [NII]$\lambda$6583 with respect to H$\alpha$. 
From the \nxmps\ galaxies in Table~\ref{main_list}, only 
31 are known as inferred
by comparison with the compilation carried out in \paperi , and with 
the new targets in \citet{2012A&A...546A.122I}. The remaining 165 
objects seem to be new members of the XMP class.
Among the unknown members, 18 have  $12+\log({\rm O/H}) \le 7.50$.

As part of the procedure to measure O/H, we also determine
the ratio N/O (Sect.~\ref{hcm}). Closed-box stellar evolution over long
timescales  
predicts $\log({\rm N/O})$ to be of the order -1.5 
at low metallicity, and when $12+\log({\rm O/H}) > 8$ it increases 
linearly with increasing metallicity reaching $\simeq -1.0$ 
at solar metallicity \citep[e.g.,][]{1999PASP..111..919H,2005A&A...434..531K}.
Most XMPs have N/O consistent with the value expected at low metallicity, however, 
a number  of them shows $\log({\rm N/O})\simeq -1.0$ or even larger 
(Figs.~\ref{distribution}a  and \ref{masmetal2}).  Metal-poor 
galaxies with $\log({\rm N/O})\simeq -1.0$ are not very common, but 
examples do exist in the literature. For instance, the so-called 
 {\em green-pea} galaxies have high N/O despite their low metallicity
\citep[][; we have  several such green peas among our XMP galaxies --  
see Sect.~\ref{morphology}]{2010ApJ...715L.128A,2012ApJ...754L..22A}. 
Moreover, some of the classical XMPs
also show enhanced N/O \citep{2014ApJ...797...81M}.
XMPs with large N/O do not appear in \paperi ,
since we selected galaxies having weak N lines and so
our search was biased toward XMPs of low N/O . 
A simple interpretation for having both large N/O and low O/H is the 
accretion of significant amounts of metal-poor gas onto 
galaxies having solar metallicity inter-stellar medium 
\citep[ISM; e.g.,][and references therein]{2010ApJ...715L.128A}. 
The mixing of metal-poor gas with pre-existing metal-rich gas 
drops down the metallicity (i.e., increases H in O/H), but it does 
not modify the original N/O, which remains as for a metal-rich ISM.
Some 10\,\%\ of the XMPs have $\log({\rm N/O}) \geq -1.2$,
representing objects where the accretion of metal-poor gas
is arguably part of their star-formation history.
Note that accretion will remain unnoticed when the pre-existing 
ISM presents low N/O.
Interpreting N/O as a tracer of previous star-formation episodes
is consistent with the trend we observe for the reddening to increase with 
increasing N/O (Fig.~\ref{extinction}b). The correlation
naturally arises if part of the dust responsible for extinction was
produced together with the N and O in previous star formation 
episodes.

We quantify the global properties of the XMPs.
They have a mean stellar mass around $10^8\,{\rm M}_\odot$,
but with very large scatter going from  $10^6\,{\rm M}_\odot$ to  
$10^{10}\,{\rm M}_\odot$, therefore some of them are fairly massive  
(Sect.~\ref{masses}).
We find that both N/O and extinction increase with increasing 
stellar mass, so that the contribution of evolved stellar populations to
the metal content and dust is more important in massive galaxies.  
From the reddening of the emission line spectrum we infer
the dust-mass column density to be around 0.2\,${\rm M_\odot\, pc^{-2}}$,
which renders a very moderate extinction coefficient (Sect.~\ref{reddening}).
For typical star-forming region size, the dust mass is between 
200 and 20,000 ${\rm M}_\odot$. 
We analyze the morphology of the XMPs, finding results in the line 
of previous studies showing a strong tendency for the XMPs to be
tadpole-like or cometary 
(\citeauthor{2008A&A...491..113P}~\citeyear{2008A&A...491..113P};
\paperi ;
\citeauthor{2013A&A...558A..18F}~\citeyear{2013A&A...558A..18F}).
Explicitly, 71\,\%\ are cometary or knotted, and only 23\,\%\ of them
are symmetric  (see Sect.~\ref{morphology}).
As a by-product of the procedure to determine gas phase metallicity,
we fit the observed continuum spectrum as a superposition of 
single stellar populations of varied ages and metallicities. According to such 
decomposition, the XMPs are characterized by a fairly young stellar 
spectrum ($< 1\,{\rm Gyr}$), although young and aged stellar populations 
co-exists at the low-metallicity starbursts (Sect.~\ref{agess}).

Since the parent sample used to select the XMPs is 
limited in apparent magnitude, it is relatively simple to
estimate the number density of XMPs in the local Universe.
We worked it out in Sect.~\ref{number_density}, finding some $3.4\times 10^{-3}$
XMPs per cubic Mpc (Eq.~[\ref{actual_density}]).
This density is significantly larger than the density 
obtained from the sample in \paperi, and it represents some 2\,\% of
all galaxies in a volume. 
We also work out the large-scale environment of the XMPs using 
constrained cosmological numerical simulations (Sect.~\ref{environment}).
XMPs have a strong tendency to appear in voids and to avoid cluster 
environments.

One of the most intriguing properties of XMPs is the existence of a
lower limit metallicity, of the order of  2\,\%\ the solar metallicity.
The limit remains in our list (Sect.~\ref{threshold}),
despite the fact that the sample is significantly larger than previous
one, and that we have systematically search the full SDSS 
spectroscopic sample. One of the ways out is, once again, 
the infall of inter-galactic gas feeding the star-formation process.  
The inter-galactic gas
collects metals ejected from galaxies along the Hubble 
time, which yields a small but non-zero metallicity expected to be 
around the value of the observed threshold 
\citep[][and references therein]{2014A&ARv..22...71S}.    

A caveat to keep in mind is that any search for XMPs based 
on optical surveys, such as the one we present, 
overlooks faint low surface brightness galaxies. 
We are sensitive to star-bursting XMPs (see Sect.~\ref{introduction}).
As pointed out 
by, e.g., \citeauthor{2013AJ....146....3S}~(\citeyear{2013AJ....146....3S}; see also 
Sect.~\ref{introduction}),
most XMPs may be low surface brightness quiescent XMPs,
so that to identify them
one would have to resort to blind HI surveys like the  
Arecibo Legacy Fast ALFA Survey \citep[ALFALFA;][]{2011AJ....142..170H}
that led to the discovery of Leo~P \citep{2013AJ....146...15G}.

\acknowledgements
Thanks are due to 
Amanda del Olmo for pinpointing repetitions in the original list of objects, 
to Andr\'es Asensio-Ramos for assistance with the estimate of false positives and 
negatives,
to Mercedes Filho for enlightening discussions and suggestions on various 
aspects of the work, 
to Steffen He{\ss} for help with the simulations used in Sect.~\ref{environment},
and 
to Ricardo Amor\'\i n and Jos\'e V\'\i lchez for a critical reading of the original manuscript.
Thanks are also due to an anonymous referee for pointing out the 
need to explicitly mention the existence of 
bursting and quiescent XMPs, 
and to Evan Skillman for fruitful discussions on the issue.
%
%
This work has been partly funded by the Spanish Ministry of Economy and 
Competitiveness, project {\em Estallidos} AYA2013--47742--C4--2--P.
RGB acknowledges support from the 
Spanish Ministerio de Ciencia e Innovación under grant AYA2010-15081.
SEN acknowledges support by the Deutsche Forschungsgemeinschaft under the grant NU 332/2-1.
Funding for SDSS, SDSS-II, and  SDSS-III has been provided by the Alfred P. Sloan Foundation, 
the Participating Institutions, the National Science Foundation, 
and the U.S. Department of Energy Office of Science.
The {\sc starlight}  project is supported by the Brazilian agencies 
CNPq, CAPES and FAPESP and by the France-Brazil CAPES/Cofecub
program.
%

%
%
\newcommand\jcap{JCAP}
 \newcommand\rmxaa{RMxAA}
\newcommand\nar{New Astron. Rev.}
\bibliographystyle{aa}

\appendix
\section{Number of false positives and negatives}\label{appendix}
For the sake of compactness, we will use the symbol $X_i$ to denote the value of 
12+log(O/H) in the i-th XMP galaxy.
By definition, $X_i < X_0$, with $X_0$ the threshold to be XMP.
Bayes's theorem \citep[e.g.,][]{martin71}
provides the probability that the true metallicity is $X$ given that
we have observed $X_i$,  
\begin{equation}
P(X| X_i)=P(X_i|X)\, H(X)/P(X_i),
\label{false1}
\end{equation}
where $P(X_i|X)$ stands for the probability of measuring $X_i$ when the true metallicity is 
$X$, $H(X)$ represents the probability the galaxy has a true metallicity $X$, and
$P(X_i)$ corresponds to the probability that the measured metallicity of the galaxy
is $X_i$ considering all possible values of the true metallicity, i.e.,
\begin{equation}
P(X_i)=\int_{-\infty}^\infty P(X_i|X)\,H(X)\,dX.
\end{equation}  
If $X$ is larger than the threshold metallicity $X_0$,
Eq.~(\ref{false1}) provides the probability that the galaxy gives a false positive 
if its true metallicity is $X$. Since the true metallicity yielding false positives
can be any value larger than the threshold, the total probability of a false positive 
$P(X \geq X_0|X_i)$ is just the sum over all possibilities, i.e.,
an integral of  Eq.~(\ref{false1}) for all $X \geq X_0$, namely, 
\begin{equation}
P(X \geq X_0 | X_i)=\int_{X_0}^\infty P(X_i|X)\,H(X)\,dX\Big/\int_{-\infty}^\infty P(X_i|X)\,H(X)\,dX.
\label{false2}
\end{equation}
Assuming the distribution of observational errors to be Gaussian, the probability of 
measuring $X_i$ when the true metallicity is $X$ turns out to be,
\begin{equation}
P(X_i|X)=\exp\big\{-{{1}\over{2}}[(X_i-X)/\sigma_i]^2\big\}/\sqrt{2\pi}\sigma_i,
\label{false3a}
\end{equation}
with $\sigma_i$ the error, i.e., the square root of the variance of the distribution of errors. 
Equation~(\ref{false2}) quantifies the probability that a single galaxy gives a false positive,
i.e., it gives the number of false positives per galaxy. When we consider the full 
set, then the total number of false positives $N_{\rm false+}$ is the sum of the number 
of false positives for all XMP galaxies, 
i.e.,
\begin{equation}
N_{\rm false+}=\sum_{i}^{\forall\, {\rm XMP}}\,P(X \geq X_0 | X_i).
\label{false4a}
\end{equation}

By definition, false negatives are true XMPs than end up having  $X_i > X_0$.
The number of false negatives, $N_{\rm false-}$,  can be 
worked out in a way similar to the derivation of $N_{\rm false+}$. This time the sum 
is over all non-XMP galaxies,
\begin{equation}
N_{\rm false-}=\sum_j^{\forall\,{\rm non-XMP}}\,P(X < X_0 | X_j),
\label{false5}
\end{equation}
with 
\begin{equation}
P(X < X_0 | X_j)=\int^{X_0}_{-\infty} P(X_j|X)\,H(X)\,dX\Big/\int_{-\infty}^\infty P(X_j|X)\,H(X)\,dX,
\label{false6}
\end{equation}
giving the probability of being XMP even though the measured metallicity $X_j$ is larger than
the threshold $X_0$. 

Assuming that the probability density function of possible oxygen abundances $H(X)$
is given by the observed histogram of abundances in  Fig.~\ref{distribution}b, and 
using the metallicities and their errors derived in Sect.~\ref{list_of_xmp}, 
Eqs.~(\ref{false2}), (\ref{false3a}), (\ref{false4a}), (\ref{false5}), and (\ref{false6}) yield,
\begin{equation}
N_{\rm false+}\simeq 82,
\label{false3}
\end{equation}
\begin{displaymath}
N_{\rm false-}\simeq 46.
\end{displaymath}
If we assume a uniform distribution of abundances in the interval $7 \leq 12+\log({\rm O/H}) \leq 8.5$, then  
\begin{equation}
N_{\rm false+}\simeq N_{\rm false-} \simeq 55.
\label{false4}
\end{equation}

\end{document}